\documentclass[11pt,draft]{article}
\usepackage{amsmath}
\usepackage{amssymb}
\usepackage{latexcad}

\newcommand{\prirodni}{\ensuremath{\mathbb{N}}}     
\newcommand{\celi}{\ensuremath{\mathbb{Z}}}         
\newcommand{\realni}{\ensuremath{\mathbb{R}}}       

\newcommand{\sgn}{\mathop{\rm sgn}\nolimits}          

\newcommand{\ds}{\displaystyle}    
\newcommand{\del}{\partial}       

\renewcommand{\leq}{\leqslant}  
\renewcommand{\geq}{\geqslant}  

\newcommand{\sestj}[6]{ \left\{ \begin{array}{ccc} #1 & #2 & #3 \\ #4 & #5 & #6
\\ \end{array} \right\}  }   

\newcommand{\trij}[6]{ \left( \begin{array}{ccc} #1 & #2 & #3 \\ #4 & #5 & #6
\\ \end{array} \right)  }   

\newcommand{\DL}{{F_{DL}}}
\newcommand{\tS}{\tilde{S}}
\newcommand{\cA}{{\cal A}}
\newcommand{\cI}{{\cal I}}
\newcommand{\cJ}{{\cal J}}
\newcommand{\cL}{{\cal L}}
\newcommand{\cN}{{\cal N}}
\newcommand{\cO}{{O}}
\newcommand{\cS}{{\cal S}}
\newcommand{\cY}{{\cal Y}}

\begin{document}
\thispagestyle{empty}
\centerline{\Large{\bf Large-spin asymptotics of Euclidean LQG}}
\centerline{\Large{\bf flat-space wavefunctions}}
\bigskip
\bigskip
\bigskip
\centerline{Aleksandar Mikovi\'c\footnote{Member of the Mathematical Physics
Group,
University of Lisbon. E-mail: amikovic@ulusofona.pt}}
\centerline{\textit{Departamento de Matem\'atica, Faculdade de Engenharias e
Ci\^encias Naturais}} \centerline{\textit{Universidade Lus\'{o}fona de
Humanidades e Tecnologia}}
\centerline{\textit{Av. do Campo Grande, 376, 1749-024, Lisboa,
Portugal}}
\bigskip
\centerline{and}
\bigskip
\centerline{Marko Vojinovi\'c\footnote{E-mail: vmarko@cii.fc.ul.pt}}
\centerline{\textit{Grupo de F\'isica Matem\'atica da Universidade de Lisboa}}
\centerline{\textit{Av. Prof. Gama Pinto, 2, 1649-003 Lisboa, Portugal}}

\bigskip
\bigskip
\begin{abstract}
We analyze the large-spin asymptotics of a class of spin-network wavefunctions
of Euclidean
Loop Quantum Gravity, which corresponds to a flat spacetime. A wavefunction from
this class can be represented as a sum over the spins of an amplitude for a spin network whose
graph is a composition of the the wavefunction spin network graph with the dual one-complex graph and the tetrahedron graphs for a triangulation of the spatial 3-manifold.
This spin-network amplitude can be represented
as a product of $6j$ symbols, which is then used to find the large-spin
asymptotics of the wavefunction. By using the Laplace method we show that the
large-spin asymptotics is given by a sum of Gaussian functions. However, these
Gaussian functions are not of the type which gives the correct graviton
propagator.
\end{abstract}

\newpage
\section{\label{SecIntroduction}Introduction}

Loop Quantum Gravity (LQG) is a theory of non-perturbative and
back\-gro\-und-independent quantization of GR, see \cite{lqg}. It is based on
the canonical quantization method, and instead of spatial metric, a spatial
connection is used as the configuration space variable. Consequently, the
Hilbert space of physical states is spanned by the spin network states
$|\Gamma\rangle$, where $\Gamma$ is a closed $SU(2)$ spin network. The graph of
$\Gamma$ is a combinatorial graph, i.e. it is a homotopy class, because the
spin-network states are diffeomorphism invariant. A physical state
$|\Psi\rangle$ is given as a linear combination of the spin-network states such
that it satisfies the quantum Hamiltonian constraint ${\mathcal H}|\Psi\rangle =
0$, where $\cal H$ is the Hamiltonian constraint operator.

Solving the Hamiltonian constraint is a difficult problem, and various
strategies have been developed over the years. In particular, one can consider
the quantum Hamiltonian constraint in the Ashtekar connection representation,
where $\cal H$ becomes a polynomial in the functional derivatives. Then any
functional $\Psi(A)$ with a support on flat connections is a solution when the
cosmological constant is zero, see \cite{mik1}. The Ashtekar connection is
complex in the Minkowski signature case so that the resolution of the identity
is given by
 \begin{equation}\int{\cal D}(Re\,A) {\cal D}(Im\,A) |A\rangle\langle A|= I
\,,\label{cri} \end{equation}
since the operator $A$ is similar to the annihilation operator for the harmonic
oscillator. The resolution of the identity (\ref{cri}) is an obstacle to
construct the loop transform from $\Psi(A)$ to the spin network wavefunction
$\Psi(\Gamma)$. However, in the Euclidean signature case, the Ashtekar
connection is real, so that one obtains the usual expression
$$\int{\cal D}A |A\rangle \langle A| = I\,.$$
Consequently
 \begin{equation}\langle \Gamma |\Psi\rangle = \int {\cal D}A \,
\overline{W_\Gamma (A)} \,\Psi (A) \,,\label{lt} \end{equation}
where $W_\Gamma (A) = \langle A |\Gamma\rangle$ is the generalization of the
trace of the holonomy along a curve to a spin network, see \cite{lqg}.

The path integral (\ref{lt}) can be rigorously defined by using a
three-di\-men\-si\-o\-nal spin-foam state sum for the quantum group $U_q
(su(2))$ where $q=\exp\left(\frac{i\pi}{k+2}\right)$ is a root of unity
\cite{mik1,mik2}. When $\Psi(A)= const \cdot\delta (F)$ then (\ref{lt}) becomes
an invariant $\Psi_k(\Gamma)$, which is proportional to the
Witten-Reshetikhin-Turaev invariant for a spin network $\Gamma$ embedded in a
compact 3-manifold $\Sigma$ representing  a spatial slice of the spacetime
\cite{marm}. This means that the  state
 \begin{equation}|\Psi_k\rangle = \sum_\Gamma \Psi_k (\Gamma) |\Gamma\rangle
\label{ks} \end{equation}
corresponds to the Kodama wavefunction
$$\Psi_k (A) = \exp\left(i\frac{k}{4\pi}\int_\Sigma Tr\left(A\wedge dA +
\frac{2}{3}A\wedge A\wedge A\right)\right)\,.$$
Therefore the effect of using a quantum $SU(2)$ group is that the
flat-con\-nec\-ti\-on state, represented by the wavefunction $\Psi(A)$, is
deformed into the Kodama state (\ref{ks}). This state is a physical state for
quantum GR with the cosmological constant $\Lambda_k$ proportional to $1/k$,
where $k$ is an
integer.

When $\Psi(A) = \exp\left(i\int_\Sigma Tr(E^m A_m)\right)\delta (F)$, it can be
argued that the corresponding quantum group state $|\Psi_k \rangle$ is a
physical state for quantum GR with the cosmological constant $\Lambda_k$
such that the triads take the values $E_m$, see \cite{mik2}. Hence the spin-network wavefunction
$\Psi_k (\Gamma, E)$ gives a physical state describing a spacetime whose spatial
metric is given by $g_{mn}=Tr(E_m E_n)$. In particular, one can choose the flat
triads and hence $|\Psi_k \rangle$ can be considered as a Euclidean analog of
the vacuum state for a De Sitter spacetime. Since $\Lambda_k \to 0$ as
$k\to\infty$, then $|\Psi_k \rangle$ for large $k$ can be considered as a good
approximation for the Euclidean flat-spacetime vacuum.

This result is very useful for the problem of finding a semiclassical
limit of LQG, since it can be shown that the graviton propagator in LQG will
have the correct large-distance asymptotics if the vacuum wavefunction $\Psi
(\Gamma, E )$ has a certain kind of Gaussian asymptotics when the spins of
$\Gamma$ are large \cite{mik3}, see also \cite{ro2,ro3} for a related approach.

As shown in \cite{mik3}, if the large-spin asymptotics is given by
 \begin{equation} \Psi (\Gamma, j_0) \approx N(\Gamma, j_0) \exp\left( -
\sum_{l,l'}\frac{C_{ll'}(\Gamma)}{j_0} (j_l-j_0)(j_{l'} -j_0)
\right)\,,\label{gas} \end{equation}
where $j_l$ is a spin of an edge $l$, $j_0$ is a parameter associated to the
flat spatial metric and $C(\Gamma)$ are $j_0$-independent and 
positive-definite matrices,
then the corresponding graviton propagator will have the correct large-distance
asymptotics.

In order to analyze the large-spin asymptotics of $\Psi_k (\Gamma,j_0)$, we will
represent it as a sum over spins of products of quantum dimensions and quantum
$6j$ symbols. When $k$ is very large, we will
approximate $\Psi_k$ by the corresponding Ponzano-Regge expression, i.e. we will
replace the quantum dimensions and the quantum $6j$
symbols in the expression for $\Psi_k$ with the corresponding classical
evaluations. The rationale for this is that $\Psi_k (\Gamma,j_0)$ was
constructed as a regularization of an expression for the zero-cosmological
constant $\Psi (\Gamma,j_0)$, which was given by the Ponzano-Regge state-sum
without the spin cut-off, see \cite{mik2}. If a spin cut-off $k/2$ is
introduced, one obtains another regularization of $\Psi(\Gamma,j_0)$, such that
the quantum group is not used. This was the regularization originally used by
Ponzano and Regge. Furthermore, the sums $\sum_0^{k/2}$ in $\Psi_k (\Gamma,j_0)$
will be replaced by the integrals $\int_0^{k/2}$, since $k$ is very large.
Consequently
 \begin{equation} \Psi (j,j_0) \approx \int_{D} d^{\omega} x \,f(x,j,j_0)
\,.\label{ntp} \end{equation}

We will split the integration region $D$ as $D=D_- \cup D_+$ where $D_+$ is the
region where all the spins are large ($x_i > j_0$), so that $\Psi \approx \Psi_-
+ \Psi_+$. Consequently the $6j$ symbols from $\Psi_+$ can be approximated by
the Ponzano-Regge formula, while the $6j$ symbols from $\Psi_-$ can be
approximated by the
asymptotic formulas for $5$, $4$ and $3$ large spins, which were also found by Ponzano and Regge, see \cite{PR}. We will study in detail the
asymptotics of $\Psi_+$, while the analysis of $\Psi_-$ will be only sketched
since it is very similar to the analysis of $\Psi_+$ and it can be shown that
the $\Psi_-$ asymptotic contribution is subleading to that of $\Psi_+$. 

We will use the Laplace method to find the asymptotics of (\ref{ntp}) for large
$j$ and $j_0$. In order to do this we will first approximate $f$ as a sum of
exponentials, see section \ref{sekcijaPreliminarnaAnaliza}. Then it is not difficult to show
that
 \begin{equation}\Psi(j,j_0) \approx \sum_{n,\pm}  N^\pm_n (j_0)
\,e^{-\frac{1}{2}(j - j_0)^T \left(B_n^\pm +\frac{1}{j_0} C_n^\pm + O(1/j_0^2 )
\right)(j - j_0)} \,,\label{gena}
\end{equation}
where $B_n^\pm$ e $C_n^\pm$ are constant ($j_0$-independent) matrices.

The expression (\ref{gena})
will give the desired asymptotics if $B_n^\pm  = 0$ for all $n$ and all spin
networks and $C_n^\pm \ne 0$ for some $n$ of some spin network. In the following
sections we will show that either $B_n^+ \neq 0$ or $C_n^+ = 0$ for all $n$ and
all spin networks. We will also show that $C_n^- = 0$ for all
$n$ and all spin networks, so that we will prove that $\Psi (j,j_0)$ does not have the desired
asymptotics.

The paper is organized in the following way. In section \ref{SecGeneral} we
briefly explain
the construction of the relevant spin-network wavefunctions from
\cite{mik1,mik2,marm}. In section
\ref{sekcijaPreliminarnaAnaliza} we outline the main procedure of analyzing the
wavefunction asymptotics, and introduce the notation. Section \ref{Sekcija4}
deals with the detailed analysis of the large-spin asymptotics of the $\Psi_+$
part of the wavefunction. The $\Psi_+$ is written as an integral of an
exponential function,
which is suitable for the stationary-point approximation. The
integration is then performed in section \ref{SecSaddlePoint} and the result is
a
sum of Gaussian functions of the form similar to (\ref{gas}), but with a
more general matrix coefficient in the exponent, denoted as $\tS$. In section
\ref{SecAsymptotics} we analyze this coefficient, and prove that it never has
the form (\ref{gas}). In order to demonstrate and verify this result further,
in section \ref{SecComputation} the matrix $\tS$ is explicitly computed for
two simple spin-networks, which are a loop spin network and a theta spin network.
In section \ref{SecPsiMinus} we discuss the large-spin asymptotics of the $\Psi_-$ part of the
wavefunction. We present our conclusions in
\ref{SecConclusion}, while in the Appendix we give all the necessary formulas and prove a
matrix theorem which determines the asymptotics of the $\tS$ matrix.

\section{\label{SecGeneral}Physical spin-network wavefunctions}

Let $\Sigma$ be a compact 3-manifold, and let $\Gamma = \{\gamma, j_l , \iota_v
\}$ be a spin network embedded in $\Sigma$, where $\gamma$ is the spin-network
graph, $j_l$ denote the edge spins and $\iota_v$ denote the vertex intertwiners.
Let $\Delta(\Sigma)$ be a
triangulation of $\Sigma$ adapted to $\gamma$ in the following way: let $H$ be a
handle-body obtained by thickening of the dual one-complex $\Delta_1^*
(\Sigma)$. The graph $\gamma$ is embedded in $H$ such that each vertex
$v_\gamma$ is placed in a different 3-handle of $H$ and each edge $l_\gamma$
runs through appropriate one-handles of $H$, see Fig 1. Let $L_H$ be the
Chain-Mail link
associated with $H$ and let $L_l$ be a set of loops associated to the edges of
$\gamma$, such that $L_l$ is embedded in the one-handle of $H$ associated to an
edge $l$ and $L_l$ is linked with the meridian loop for that one-handle, see Fig
2.

\begin{center}
\setlength{\unitlength}{0.5mm}

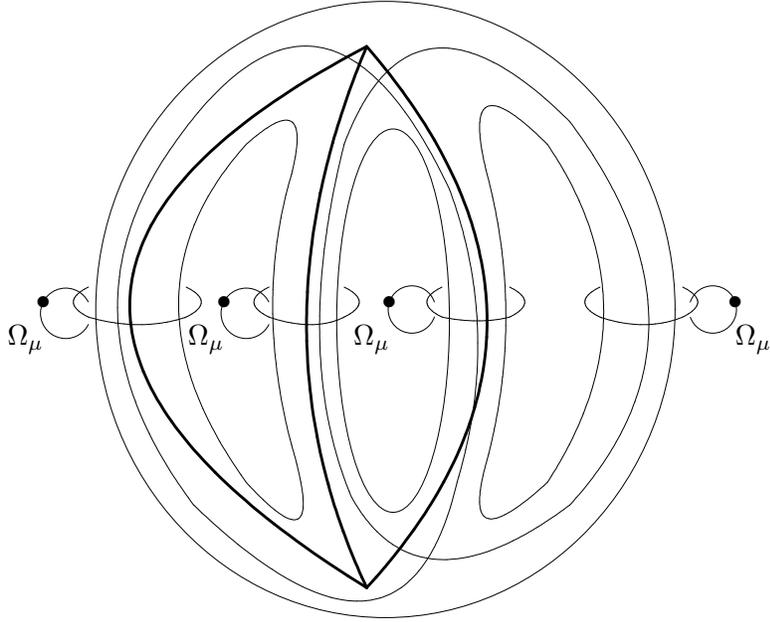
\\ Figure 2.
\end{center}

Let $\phi: \Sigma \to S^3$ be a smooth map from $\Sigma$ to a 3-sphere, and let
us color the link $L_H$ with the $\Omega$ elements. The $\Omega$ element is a
linear combination of colors given by
$$\Omega = \sum_{j=0}^{k/2} \dim_q j \,C(j)\,,$$
where $C(j)$ denotes the color (spin $j$) associated to a loop of $L_H$. We also introduce
$$\Omega_{\mu} = \sum_{j=0}^{k/2} \mu(j) \,C(j)\, .$$

Let us color the $L_l$ loops with $U_q (su(2))$ irreps $\lambda_l$. We will
denote
the quantum group evaluation of the colored link $L_H \cup L_1 \cup \cdots\cup
L_n \cup \gamma$ embedded in $S^3$ as
\begin{equation}\langle L_H \cup L_1 \cup \cdots\cup L_{n_2} \cup
\gamma,\Omega^{n_2 +n_3}, j,\iota,\lambda\rangle \,,\label{cmle}\end{equation}
where $n_2$ is the number of dual edges (triangles) and $n_3$ is the number of
dual vertices (tetrahedrons).

From the properties of the $\Omega$ element, see \cite{marm}, it follows that
the evaluation (\ref{cmle}) can be expressed as
$$ \sum_{j',\iota'} \prod_f \dim_q j'_f \, \langle \Delta_1^* \cdot
(Tet)^{n_3}\cdot\gamma, j',\iota',j,\iota,\lambda\rangle\,,$$
where $\{\Delta_1^* \cdot (Tet)^{n_3}\cdot\gamma, j',\iota',j,\iota,\lambda\}$
is the spin network which is obtained by removing $n_2$ $\Omega$-elements from
$n_2$ 1-handle meridians.

Then
\begin{equation} \Psi_k (\Gamma, E) = \sum_{\lambda,j',\iota'}\prod_l
\mu(\lambda_l,E_l)\prod_f \dim_q j'_f \, \langle \Delta_1^* \cdot
(Tet)^{n_3}\cdot\gamma, j',\iota',j,\iota,\lambda\rangle\,,\label{vwf}
\end{equation}
where
$$ \mu (\lambda, E ) = \frac{1}{\dim\lambda}\int_{SU(2)} dg
f(g,E)\,\chi^{(\lambda)} (g) \,,$$
and $\chi_\lambda$ is the trace of the $\lambda$-representation matrix of
$g$. The function $f(g,E)$ is determined by the choice of the flat-connection
wavefunctional given by
$$ \Psi(A) = \exp\left(i\int_\Sigma Tr(E^m A_m)\right)\psi(A)\delta (F)\,.$$
This $\Psi(A)$ solves the Hamiltonian constraint for $\Lambda = 0$, but if
$\psi(A) \ne 1$, then $\Psi(A)$ does not have the sharp values for the tetrads,
which is the LQG equivalent of replacing a plane-wave with a coherent state.

One can introduce the background spins $j^0_l$ associated with the triads $E$,
via the relation $|E_l| = j^0_l L^2$, where
$$E_l = \int_{\Delta_l} E^m \,\epsilon_{mnp} \, dx^n \wedge dx^p \,,$$
$\Delta_l$ is the triangle dual to a dual edge $l$ and $L$ is the Planck
length.
In the flat-triad case, one can assume that $|E_l|=const$, and therefore $j_l^0
= j_0$, so that
$ \Psi_k (\Gamma, E)= \Psi_k (\Gamma, j_0)$. We will then choose
\begin{equation}\mu(\lambda) = \frac{e^{-(\lambda-j_0)^2}}{2\lambda+1}\,,\label{gamu}\end{equation}
in order to mimic the Rovelli ansatz for the
wavefunction \cite{ro2}. We will then examine the asymptotics of (\ref{vwf}) for
large spins $j$ and $j_0$.

We will express the evaluation of the $\Delta_1^* \cdot (Tet)^{n_3}\cdot\gamma$
spin network as a sum of products of quantum $6j$ symbols, since this will
facilitate our analysis of the large-spin asymptotics. This can be done because
the evaluation of an arbitrary three-valent spin network $\Gamma$ can be
represented as a sum of products of $6j$ symbols, see \cite{T,marm}.

Let $\Gamma'$ be a projection of $\Gamma$ onto an $S^2$. The graph $\gamma'$
will divide the sphere into disjoint discs. Color the discs with $SU(2)$ irreps
$\alpha_1,..., \alpha_{n}$ and write the corresponding Turaev shadow-world
evaluation
$w (\Gamma,\alpha)$, which is given as a product of $6j$ symbols. Then
 \begin{equation} \langle \Gamma \rangle \propto \sum_{\alpha}\prod_{i=1}^n
\dim_q \alpha_i \,w (\Gamma,\alpha)\,.\label{sjr} \end{equation}
Alternatively, (\ref{sjr}) is the quantum group evaluation of the link formed by
$\gamma'$ and the Chain-Mail link for a two-dimensional handle-body which is a
thickening of $\gamma'$ \cite{marm}.

By using (\ref{vwf}) and (\ref{sjr}) we obtain the following expression for the
spin-network wavefunction
\begin{equation}
 \Psi_k (j,j_0) = \sum_{a,\alpha,\lambda,\iota} \left(
\prod_a d_a \prod_{\alpha}d_{\alpha}\prod_{\lambda} \mu(\lambda) \prod_v \{6j_v
\}  \right) \,,\label{sixjr}
\end{equation}
where
\begin{itemize}
\item spins $j$ label the graph $\gamma$,
\item spins $a$ label the faces of $\Delta^*$,
\item spins $\lambda$ label the edges of $\Delta^*_1$,
\item spins $\alpha$ label the disjoint discs of the projected graph $\Delta_1^*
\cdot (Tet)^n \cdot \gamma$,
\item $\{6j_v \}$ denotes the $6j$-symbol associated to a vertex $v$ of the
projected graph $\Delta_1^* \cdot (Tet)^{n_3} \cdot \gamma$,
\item $d_x$ is the quantum dimension of the representation $x$.
\item  $\iota$ denote the intertwiners
for the graph $\Delta_1^* \cdot (Tet)^{n_3} \cdot \gamma$.
\end{itemize}

For example, when $\Gamma =\{\gamma,j\}$ where $\gamma$ is a loop embedded in
$\Sigma = S^3$, we can triangulate $S^3$ with two tetrahedrons such that
$\Delta_1^*$ is a theta-four graph $\theta_4$. The corresponding $3$-valent
$\theta_4 \cdot (Tet)^2 \cdot \gamma$ spin network is given in Fig 3. in the
Appendix.

We will also use the Latin indices $M,N,p,q,r,s$ to denote any of the
$j,a,\alpha,\lambda$ and $\iota$ indices, and
$$
j \in \cJ = \{ 1,\dots,J \}, \qquad a\in \cA= \{ J+1,\dots,J+A \},
$$
$$
\alpha, \iota \in \cY= \{ J+A+1,\dots J+A+\Upsilon \},
$$
$$
\lambda \in \cL= \{ J+A+\Upsilon+1,\dots, J+A+\Upsilon+L \},
$$
$$
N \in \cN= \{ 1, \dots, \Omega \}, \qquad \Omega \equiv J+A+\Upsilon+L.
$$
The corresponding sums and products will be over the whole domain appropriate
for each type of index, unless otherwise noted. We will often have the set of
all values of the
indices except for the $j$-indices, so we denote it as
$$
\cI = \cN / \cJ.
$$
It has $\omega \equiv \Omega - J$ elements.

The $6j$ symbols $\{6j_v \}$ are functions of all spins from a set $J$. The
index $v$ will also
be treated as a multi-index for spins, in the sense that the $6j$ symbol
$$
\sestj{x_1}{x_2}{x_3}{x_4}{x_5}{x_6}
$$
is enumerated by the index $v=(1,2,3,4,5,6)$, representing the ordered $6$-tuple
of spins which appear in that $6j$ symbol. This will be useful in limiting the
domain of indices to a particular $6j$ symbol, noted as $M,N\in v$.

The function $\mu(\lambda)$ can be chosen arbitrarily, but as we explained in
section \ref{SecGeneral}, we will choose the Gaussian function (\ref{gamu}).
In this way a spin scale, $j_0$, is conveniently introduced in the
wavefunction.

\section{\label{sekcijaPreliminarnaAnaliza}Preliminary analysis}

As we explained in the introduction, the quantum group expression (\ref{sixjr})
will be replaced by the corresponding classical group expression, where the
spins will have a cut-off given by $k/2$. Since we are interested in the case
where $k$ is a very large number, then the finite sums $\sum_0^{k/2}$ in
(\ref{vwf}) can be
approximated by the integrals $\int_0^{k/2}$, so that
 \begin{equation} \Psi (j,j_0) \approx \int_{D} d^{\omega} x \,f(x,j,j_0)
\,,\label{intp} \end{equation}
where
$$
f (x,j,j_0) = \prod_a
(2a+1) \prod_\alpha (2\alpha + 1)\prod_{\lambda} \frac{e^{-(\lambda
-j_0)^2}}{2\lambda + 1} \prod_v \{ 6j_v \}.
$$
The domain $D$ is a subset of $(k/2)^N$ and $D$ is determined by the triangle
conditions for
the spins coming from the $6j$ symbols.

Let us split the integration region $D$ as $D = D_- \cup D_+$ where $D_+$ is the
region where all the spins are large ($x \ge j_0$ for every $x$), and $D_- =
D\setminus D_+$. Consequently
\begin{equation} \label{PsiPlusPsiMinus}
\Psi(j,j_0) \approx \int_{D_-} d^{\omega} x \,f(x,j,j_0) + \int_{D_+} d^{\omega}
x
\,f(x,j,j_0)= \Psi_- (j,j_0)+ \Psi_+ (j,j_0) \,.
\end{equation}
The $6j$ symbols from $\Psi_+$ can be approximated by the PR formula, while the
$6j$ symbols from $\Psi_-$ can be approximated by the asymptotic formulas for
$5$, $4$ and $3$ large spins. Namely, there will be a
certain number of tetrahedrons which contain the large spins $j$, so that each
of these tetrahedrons will have at least two other large spins, due to the
triangle inequalities. Each of these large spins appear in other tetrahedrons,
which will force another spins to be large, and so on. In the end there will be
a substantial number of large spins different from $j$, but some of the internal
spins can still remain small\footnote{In the case of the loop spin network, a
numerical investigation has given $11.000$ configurations with small spins.
The
maximal number of small spins was 9 out of 22 spins.}. Consequently
$$ \Psi_\pm (j,j_0) \approx \int_{D_\pm } d^{\omega} x \,f_\pm (x,j,j_0) \,,$$
where $f_\pm$ are the corresponding approximations for $f$ in $D_\pm$ regions.

We will use the Laplace method to find the asymptotics of (\ref{intp}) for large
$j$ and $j_0$. In order to do this we will first approximate $f_\pm$ as sums of
exponentials. Namely, if $x_n$ are the stationary points of $f(x)$ then
 \begin{equation} f(x,j,j_0) \approx \sum_n \epsilon_n e^{-S_n
(x,j,j_0)}\,,\label{gaex} \end{equation}
where $S_n (x,j,j_0) \approx |\ln |f(x,j,j_0)||$ in the vicinity of $x_n$ and
$\epsilon_n = \pm 1$ depending on whether $x_n$ is a minimum or a maximum.
Consequently
$$\Psi (j,j_0) \approx \sum_n \epsilon_n \int_{D} d^N x  \, e^{-S_n (x,j,j_0)} =
\sum_n \epsilon_n \, I_n\,,$$
and one can apply the Laplace method to evaluate the integrals $I_n$. This gives
 \begin{equation}\Psi \approx \sum_n \epsilon_n \sum_{x^*,j^*} N_n^* (j_0)
\,e^{-\frac{1}{2}(j - j^*_n)^T \tilde{S}_n^* (j_0)(j - j^*_n)} \,,\label{stpa}
\end{equation}
where $x^*$, $j^*$ are the stationary points of $S$ and
$$\tilde{S}^* = S_{jj}^* - (S_{xj}^* )^T S_{xx}^* S_{xj}^* $$
where $S_{jj}^*$, $S_{xj}^*$ and $S_{xx}^*$ denote the Hessian matrices in the
respective stationary points.

In order to find the matrix functions $\tilde{S}_n^* (j_0)$ we will use the
scaling
properties of $S_n$ when the spins $x,j$ and $j_0$ are scaled. Namely, let us
assume that $S_n$  satisfy
 \begin{equation} S_n (\Lambda x,\Lambda j,\Lambda j_0) = \Lambda^2 \left[ R_n
(x,j,j_0)+ O(1/\Lambda)\right] \,.\label{ssc} \end{equation}
The scaling (\ref{ssc}) will be consistent with the approximation (\ref{stpa})
if the
stationary points have the form
 \begin{equation}x^* = \mu_1^* j_0 + \mu_0^* + O( j_0^{-1})\,,\quad  j^* =
 \nu^*_1 j_0 + \nu_0^* + O( j_0^{-1})\,.\label{asol} \end{equation}

The form (\ref{asol}) of the stationary points, together with the scaling
(\ref{ssc}) and the approximation (\ref{stpa}) imply
 \begin{equation}\tilde{S}^* = B^* + C^* j_0^{-1} + O( j_0^{-2}) \,,\label{asa}
\end{equation}
where $B^*$ and $C^*$ are constant (independent of $j_0$) matrices. If $S_n$ are
such that $\nu^*_1 =1$, which will be imposed by the choice (\ref{gamu}), then
we will obtain the asymptotics  of the same type as (\ref{gas}). However, in
order to make a final comparison we need to calculate the matrices $B^*$ and
$C^*$.

\section{\label{Sekcija4}$\Psi_+$ integral }

Let us now make a more detailed analysis of the $\Psi_+$ integral. In order to
calculate the matrices $A^*$ and $B^*$ it will be convenient to introduce the
scaling parameter $\Lambda$ into the integral $\Psi_+$ through the following
change of variables
 \begin{equation}
x + \frac{1}{2} = \Lambda y \,,\quad j+\frac{1}{2}= \Lambda y_j\,, \quad j_0
+\frac{1}{2}= \Lambda y_0\,.
\label{lsc} \end{equation}
Since $x=O(\Lambda)$ and $y=O(1)$, the integration domain
$D_+$ is then transformed into $D_+'$ which is of $O(1)$. $\Lambda$ is
essentially the same as $j_0$, since $y_0 = O(1)$ and one can choose $y_0 =1$.
The
Jacobian of the transformation is $\Lambda^{\omega}$, while the factors $2a+1$
become
$2\Lambda y_a$. The spins $j$ and $j_0$ are changed into new variables $y_j$ i
$y_0$,
by using the formula (\ref{lsc}). The exponents $e^{-(\lambda -j_0)^2}$ become
$e^{-\Lambda^2 (y_{\lambda} - y_0)^2}$, so that we obtain
 \begin{equation}
\Psi_+ (j,j_0)\approx 2^A \Lambda^{A+\omega} \int_{D_+'}
d^{\omega}y \; \prod_a y_a \prod_\alpha y_\alpha \prod_{\lambda}
\frac{e^{-\Lambda^2 (y_{\lambda}-y_0)^2}}{y_\lambda}
\prod_v \{ 6j_v(\Lambda y) \}\,.
\label{lambint} \end{equation}

Since every $6j$ symbol in (\ref{lambint}) can now be approximated by the PR
formula (\ref{AsimptotikaSestJ}), we obtain
 \begin{equation} \label{TalasnaFunkcijaPrekoKosinusa}
\Psi_+ (j,j_0) \approx \frac{2^A}{(\sqrt{12\pi})^V}
\Lambda^{A+\omega-\frac{3V}{2}} \int_{D_+'} d^{\omega}y \; \prod_a y_a
\prod_\alpha y_\alpha
\prod_{\lambda} \frac{e^{-\Lambda^2 (y_{\lambda}-y_0)^2}}{y_\lambda} \prod_v
\frac{\cos\left(\cS_v(\Lambda,y)\right)}{\sqrt{V_v(y)}}\,,
 \end{equation}
where
 \begin{equation} \label{PonzanoReggeDejstvoSaPopravkomDrugo}
\cS_v (\Lambda, y) = \Lambda \sum_{s\in v} y_s \theta_{s,v}(y) + \frac{\pi}{4} +
\frac{1}{\Lambda}\DL(y) + O\left(\frac{1}{\Lambda^2} \right) \,.
 \end{equation}
The $O(\Lambda)$ term represents the Regge action, while the explicit form of
the complicated $O(1/\Lambda)$ term can
be found in \cite{Dupuis2009}.

In order to apply Laplace's method, it is vital to
rewrite the integrand as an exponential function, with the multiplicative factor
$\Lambda$ in the exponent, and to determine the positions of the extremal
points.
The main problem is that the integrand is a product of cosine functions, which
cannot be easily
cast into an exponential form. This problem can be solved by using the
approximation formula
(\ref{KosinusnaFormula}) derived in Appendix \ref{DodatakKosinusnaFormula}:
 \begin{equation} \label{AproksimacijaKosinusneFunkcije}
\cos x \approx \frac{1}{\vartheta_4(0,e^{-\frac{\pi^2}{2}})} \sum_{p\in \celi}
(-1)^p e^{-\frac{1}{2} (x-p\pi)^2}\,.
 \end{equation}

This approximation has two main advantages.
First, we avoid having to deal with complex-valued exponents, which would have
been
inevitable if we had employed the formula $\cos x = (e^{ix}+e^{-ix})/2$ and the
corresponding stationary-phase method. Using the stationary-phase method would
make the asymptotic analysis more complicated, because the corresponding
stationary points will have the coordinates which are complex numbers. Second,
we can
``capture'' the neighborhood of all extremal points {\em at once}, including
those far away --- when $p \sim O(\Lambda)$ --- in a manifest manner.
Specifically, every maximum and minimum at infinity can be labeled as $p=\Lambda
m + n$, where $m,n\in\celi$ are of the order $O(1)$.

By applying (\ref{AproksimacijaKosinusneFunkcije}) to
(\ref{TalasnaFunkcijaPrekoKosinusa}), we have
 \begin{equation} \label{IzrazZaItiKosinus}
\cos \cS_v (\Lambda,y) = \frac{1}{\vartheta_4(0,e^{-\frac{\pi^2}{2}})}
\sum_{\substack{m_v,n_v\in \celi \\ m_v,n_v \ll \Lambda }} (-1)^{\Lambda
m_v+n_v} e^{-\frac{1}{2} (\cS_v -\Lambda m_v\pi - n_v\pi)^2}.
 \end{equation}
Here it is crucial to note that the approximation is valid iff the exponent goes
to zero, which will happen in the vicinity of extremal points $y_0^*$. We
calculate the exact positions of these points by using the ansatz
 \begin{equation} \label{FundamentalniAnsatz}
y_N^* = A_N + \frac{B_N}{\Lambda} + \frac{C_N}{\Lambda^2} +
\cO\left(\frac{1}{\Lambda^3} \right).
 \end{equation}
Substituting this into the exponent and using
(\ref{PonzanoReggeDejstvoSaPopravkomDrugo}), we obtain the necessary
conditions\footnote{Here we can also note the following detail. In principle, we
could have written the integer $p$ from equation
(\ref{AproksimacijaKosinusneFunkcije}) in the more general form, which includes
some higher power of $\Lambda$, like $p= \Lambda^2 l + \Lambda m + n$. In that
case we would obtain a consistency condition $l=0$ in addition to equations
(\ref{JnaZaStacionarneTackeA}) and (\ref{JnaZaStacionarneTackeB}). Therefore,
our choice $p= \Lambda m + n$ is actually the most general nontrivial one,
dictated by the linear $\Lambda$-dependence in
(\ref{PonzanoReggeDejstvoSaPopravkomDrugo}).}
for the extremal points
$y_0^*$, in the form of two systems of equations for coefficients $A_N$ and
$B_N$
 \begin{equation} \label{JnaZaStacionarneTackeA}
\sum_{s\in v} A_s \theta_{s,v}(A) = m_v \pi,
 \end{equation}
 \begin{equation} \label{JnaZaStacionarneTackeB}
\sum_{s\in v } B_s \theta_{s,v}(A) = n_v\pi - \frac{\pi}{4}.
 \end{equation}
Of course, these equations are not a sufficient condition to determine the
extremal points, simply because we are yet to discuss the full integrand in
(\ref{TalasnaFunkcijaPrekoKosinusa}). The additional missing equations will be
determined later.

Returning now to (\ref{IzrazZaItiKosinus}), we will use
(\ref{PonzanoReggeDejstvoSaPopravkomDrugo}) to expand the exponent in
(\ref{IzrazZaItiKosinus}) in powers of
$1/\Lambda$ in order to extract the leading $\Lambda^2$ term, as needed for the
saddle-point method
$$
-\Lambda^2 \left[ \frac{1}{2} \left( \sum_{s\in v} y_s \theta_{s,v} - m_v\pi
\right)^2 + \frac{1}{\Lambda}\left( \sum_{s\in v} y_s \theta_{s,v} -m_v\pi
\right) \left( \frac{\pi}{4} - n_v\pi \right) + \right.
$$
$$
\left.
+\frac{1}{2\Lambda^2}\left( \frac{\pi}{4} - n_v\pi \right)^2 +
\frac{1}{\Lambda^2} \left( \sum_{s\in v} y_s \theta_{s,v} -m_v\pi \right) \DL +
\cO\left(\frac{1}{\Lambda^3} \right) \right].
$$
This can be done for all cosine functions in
(\ref{TalasnaFunkcijaPrekoKosinusa}), so in
the end we obtain
 \begin{equation} \label{TalasnaFunkcijaPrekoEksponenata}
\Psi_+ (j, j_0)\approx
\frac{2^A\Lambda^{A+\omega-\frac{3V}{2}}}{\left[\sqrt{12\pi}
\,\vartheta_4(0,e^{-\frac{\pi^2}{2}})\right]^V}
\sum_{\substack{m_1,n_1\in \celi \\ m_1,n_1 \ll \Lambda }} \dots
\sum_{\substack{m_V,n_V\in \celi \\ m_V,n_V \ll \Lambda }} (-1)^{\sum_v (\Lambda
m_v+n_v )}
\int_{D_+'} d^{\omega}y \; e^{\Lambda^2 S(\Lambda,y)}.
 \end{equation}
This expression can be explicitly integrated via the Laplace method, term by
term.

Here the phase has the general form
 \begin{equation} \label{RazvojMatriceSpoLambda}
S(\Lambda, y) \equiv  S_0(y) + \frac{1}{\Lambda} S_1(y) +
\frac{1}{\Lambda^2}S_2(y) + \cO\left(\frac{1}{\Lambda^3}\right),
 \end{equation}
and we have explicitly
 \begin{equation} \label{KomponenteMatriceS}
\begin{array}{ccl}
S_0(y) & = & \ds -C \sum_{\lambda} (y_{\lambda}-y_0)^2 - \frac{1}{2} \sum_v
\left( \sum_{s\in v} y_s \theta_{s,v} - m_v\pi \right)^2, \\
S_1(y) & = & \ds - \sum_v \left( \sum_{s\in v} y_s \theta_{s,v} - m_v\pi
\right)\left( \frac{\pi}{4} - n_v\pi \right), \\
S_2(y) & = & \ds \sum_a \ln y_a - \frac{1}{2} \sum_v \left[ \ln V_v +
\left( \frac{\pi}{4} - n_v\pi \right)^2 + 2 \left( \sum_{s\in v} y_s
\theta_{s,v} -m_v\pi \right) \DL \right]. \\
\end{array}
 \end{equation}
The major gain here lies in the fact that there is a systematic expansion of the
phase in powers of $1/\Lambda$, while the leading term is of the order $O(1)$.
As it will turn out, this will become very important as we go on to study the
asymptotic behavior of the whole wavefunction. We shall systematically calculate
everything up to terms of the order $O(1/\Lambda^3)$, since this is the lowest
self-consistent approximation for the Laplace method, as we shall see
below. This is also the reason why we keep the $O(1/\Lambda)$ term in
(\ref{PonzanoReggeDejstvoSaPopravkomDrugo}).

\section{\label{SecSaddlePoint}Stationary-point approximation}

We have written the wavefunction (\ref{TalasnaFunkcijaPrekoEksponenata}) in the
form
required for the application of the Laplace method. However, given that the
integral is multidimensional, there are certain complications. The first step is
to expand the phase into a power series around an extremal point $y_N^*$. We
have to cast the series in the form which separates
the variables $y_j$, which are not to be integrated over, from the internal
$y_N$ ($N \in \cI$) variables
$$
S(\Lambda,y) = S(y_n^*,y_j^*) + \sum_{N\in\cI}\frac{\del S}{\del y_N} \left( y_N
- y_N^* \right) + \sum_j \frac{\del S}{\del y_j} \left( y_j - y_j^* \right)
$$
$$
+ \frac{1}{2} \sum_{M,N\in \cI} \frac{\del^2 S}{\del y_M \del y_N} \left( y_M -
y_M^* \right) \left( y_N - y_N^* \right)
+ \sum_{\substack{ M\in\cI \\ j }} \frac{\del^2 S}{\del y_M \del y_j} \left( y_M
- y_M^* \right) \left( y_j - y_j^* \right)
$$
$$
+ \frac{1}{2} \sum_{j,j'} \frac{\del^2
S}{\del y_j \del y_{j'}} \left( y_j - y_j^* \right) \left( y_{j'} - y^*_{j'}
\right)  +\dots
$$

Given that all derivatives above are evaluated at an extremal point $y_N^*$,
the terms with first derivatives vanish. Also, since all the differences
$y_N-y_N^*$ go to zero as $1/\Lambda$, due to equations
(\ref{JnaZaStacionarneTackeA}) and (\ref{JnaZaStacionarneTackeB}), we can
neglect the terms of the order $(y-y^*)^3$ and higher. This leaves us with
$$
S(\Lambda,y) = S^* +\frac{1}{2} (y - y^*)^T S'' (y - y^*) + (y_j-y_j^*)^T
\dot{S}' (y - y^*) + \frac{1}{2} (y_j-y_j^*)^T \ddot{S} (y_j - y_j^*),
$$
where we have introduced a shorter matrix notation,
$$
S^* = S(\Lambda,y^*), \qquad
S'' = \frac{\del^2 S}{\del y_M \del y_N} \Big|_{y^*},
$$
$$
\dot{S}' =
\frac{\del^2 S}{\del y_j \del y_M} \Big|_{y^*} ,  \qquad
\ddot{S} = \frac{\del^2
S}{\del y_j \del y_{j'}} \Big|_{y^*}, \qquad (M,N\in \cI).
$$
The matrices $S''$, $\dot{S}'$ and $\ddot{S}$ are of the type
$\omega\times\omega$,
$J \times \omega$ and $J\times J$, respectively. At this point we see that
keeping all terms of the order up to $O(1/\Lambda^3)$ is necessary, since if we
had kept only terms up to $O(1/\Lambda^2)$, the phase would have been
approximated
by a constant, and the Laplace method would have not worked.

Now the integrals in the wavefunction obtain the form
$$
\begin{array}{ccl}
I & = & \ds \int_{D_+'}d^{\omega}y \; e^{\Lambda^2 S(\Lambda,y)} \\
& = & \ds e^{\Lambda^2 S^*} e^{ \frac{\Lambda^2}{2} \left( y_j - y_j^* \right)^T
\ddot{S} \left( y_j - y_j^* \right)} \int_{D_+'} d^{\omega}y \;
e^{ \frac{\Lambda^2}{2} (y - y^*)^T S'' (y - y^*) + \Lambda^2 (y_j-y_j^*)^T
\dot{S}' (y - y^*) }. \\
\end{array}
$$
At this step it is important to note that we have one integral of this type for
every
allowed value of $m_v$ and $n_v$, and for every extremal point $y_N^*$ which is
inside the integration domain $D_+'$. The Gaussian form of all the terms in
(\ref{IzrazZaItiKosinus}) guarantees that all extremal points of the integrand
are maxima, which in turn means that all eigenvalues of the matrix $S''$ are
negative or zero.

This allows us to expand the integration domain $D_+'$ to $\realni^{\omega}$,
since the eventual ``exterior'' extremal points are not taken into account while
everything else is negligible in the limit $\Lambda\to\infty$. The zero
eigenvalues contribute with linearly divergent terms, but this can be
regularized in the sense of the generalized Gaussian integral
(\ref{GeneralisaniGausovIntegral}) (see Appendix \ref{DodatakGausoviIntegrali}).
Even when the integration domain is extended to $\realni^{\omega}$
the integral converges (or has a constant divergent contribution), and the
integration can be explicitly performed by using the formula
(\ref{GeneralisaniGausovIntegral}).

Therefore, after a suitable orthogonal change of variables $z=O(y-y^*)$ which
brings $S''$ and $\dot{S}'$ in a block-diagonal form, we perform the integration
and obtain
$$
I = \sum_{y^*\in D_+'} e^{\Lambda^2 S^*} e^{ \frac{\Lambda^2}{2} \left( y_j -
y_j^* \right)^T \ddot{S} \left( y_j - y_j^* \right)} \left[ \int_{\realni} dz
\right]^{\omega-r} \frac{1}{\Lambda^r}\sqrt{\frac{(2\pi)^r}{|\det M|}}
e^{-\frac{\Lambda^2}{2} (y_j-y_j^*)^T N M^{-1}N^T (y_j-y_j^*)}.
$$
Here $r$ is the rank of matrix $S''$, while $M=M_{S''}$ (as defined in Appendix
\ref{DodatakGausoviIntegrali}). It is important to note that the
orthogonal change of basis which brings $S''$ into a block-diagonal form does
not
necessarily guarantee that $\dot{S}'$ will also reduce to zero in the
null-space of $S''$. In other words, if $K$ denotes the null-space projector of
$S''$, it is not guaranteed that $\dot{S}'K =0$, which was assumed in the above
equation. However, if this assumption is violated, we can immediately conclude
that $\Psi_+$ {\em does not have the desired asymptotics}, because of
the integral of a linear exponential function. Nevertheless, numerical
investigations (see subsections \ref{SubsekcijaVodeciRed} and
\ref{SubsekcijaPrviPopravljeniRed}) suggest that $\dot{S}'K $ is indeed zero,
so we proceed assuming that this is satisfied.

Using the fact that $\Lambda (y_j-y_j^*) = x_j-j^*$ we can switch back to
the variables $j$ and $j_0$, and write the resulting
integral as
$$
I = \sum_{y^*\in D_+'} \frac{e^{\Lambda^2 S^*}}{\Lambda^r}
\sqrt{\frac{(2\pi)^r}{|\det M|}} \left[ \int_{\realni} dz \right]^{\omega-r}
e^{ - \left( x - j^* \right)^T \tS \left( x - j^* \right)},
$$
where we have introduced the matrix
 \begin{equation} \label{JednacinaZaTildaS}
\tS \equiv - \frac{1}{2} \left( \ddot{S} - N M^{-1}N^T   \right).
 \end{equation}
This matrix is of type $J\times J$, and represents the key ingredient of the
calculation. It is known as the {\em Schur complement} and is well-studied in
general matrix theory (see Ref. \cite{Zhang2005} and Appendix
\ref{DodatakMatricneTeoreme}).

Finally, we substitute this result back in the equation
(\ref{TalasnaFunkcijaPrekoEksponenata}), and obtain
 \begin{equation} \label{IntegraljenaTalasnaFunkcijaPrekoEksponenata}
\Psi_+ (j,j_0)\approx
\frac{2^A\Lambda^{A+\omega-\frac{3V}{2}}}{\left[\sqrt{12\pi}
\,\vartheta_4(0,e^{-\frac{\pi^2}{2}})\right]^V}
\hphantom{mmmmmmmmmmmmmmmmmmmmmmm}
 \end{equation}
$$
\sum_{\substack{m_1,n_1\in \celi \\ m_1,n_1 \ll \Lambda }} \dots
\sum_{\substack{m_V,n_V\in \celi \\ m_V,n_V \ll \Lambda }} \sum_{y^*\in D_+'}
(-1)^{\sum_v (\Lambda m_v+n_v )}
\frac{e^{\Lambda^2 S^*}}{\Lambda^r} \sqrt{\frac{(2\pi)^r}{|\det M|}} \left[
\int_{\realni} dz \right]^{\omega-r}  e^{ - \left( j - j^* \right)^T \tS \left(
j - j^* \right)}.
$$

The matrices $M$, $N$ and their rank $r$ typically depend on the initial choices
of parameters $m_v$, $n_v$ and $y_N^*\in D_+'$. Due to the nature of the
Laplace method, we should keep in the sum only those parameters which
give minimum $r$, and among those only the ones which give maximum $S^*$. The
resulting sum of the remaining Gaussian functions represents the asymptotic
behavior
of $\Psi_+$.

\section{\label{SecAsymptotics}Asymptotic behavior of the matrix $\tS$}

Let us now return to (\ref{TalasnaFunkcijaPrekoEksponenata}),
$$
\Psi_+ (j,j_0)\approx
\frac{2^A\Lambda^{A+\omega-\frac{3V}{2}}}{\left[\sqrt{12\pi}
\,\vartheta_4(0,e^{-\frac{\pi^2}{2}})\right]^V}
\sum_{\substack{m_1,n_1\in \celi \\ m_1,n_1 \ll \Lambda }} \dots
\sum_{\substack{m_V,n_V\in \celi \\ m_V,n_V \ll \Lambda }} (-1)^{\sum_v (\Lambda
m_v+n_v )}
\int_{D_+'} d^{\omega}y \; e^{\Lambda^2 S(\Lambda,y)}.
$$
The phase $S(\Lambda,y)$ can be expanded into a power series around an extremal
point $y_N^*$ (the first derivatives vanish, while the third and higher-order
derivatives are of
$O(1/\Lambda^3)$), so that
$$
S(\Lambda, y) = S(\Lambda,y^*) + \frac{1}{2} (y-y^*)^T \Delta (y-y^*).
$$
Here the $\Delta$ matrix is defined as
$$
\Delta_{MN} \equiv \frac{\del^2 S(\Lambda, y)}{\del y_M \del y_N} \Big|_{y=y^*}.
$$
It is of type $\Omega\times\Omega$, and it is convenient because it can be
decomposed
into blocks of size $J$ and $\omega$
$$
\Delta \equiv [\Delta_{MN}] = \left[
\begin{array}{c|ccc}
 \ddot{S}\vphantom{\ds\int_A^A} & & \dot{S}' & \vphantom{\ds\sum} \\  \hline
          & &              &                    \\
 (\dot{S}')^T & & S''      &                    \\
          & &              &                    \\
\end{array}
\right] .
$$

After an orthogonal transformation of the basis, the matrices $S''$ and
$\dot{S}'$ will reduce simultaneously into a block-diagonal form, so that the
$\Delta$
matrix will obtain the following form
 \begin{equation}
\Delta = \left[
\begin{array}{c|cc}
\ddot{S} & \vphantom{\ds\sum}N & 0 \\ \hline
N^T \vphantom{\ds\sum}     & M                     & 0 \\
0        & 0                     & 0 \\
\end{array}
\right].
\label{deltam} \end{equation}
Integration over the zeroes in (\ref{deltam}) will boil down to a trivial
divergent part, as seen in
(\ref{IntegraljenaTalasnaFunkcijaPrekoEksponenata}), and we can
consider only the nonzero block. Note that $\tS$ matrix defined by
(\ref{JednacinaZaTildaS}) is actually (minus one half of) the Schur
complement of the nonzero block of the $\Delta$ matrix.

At this point we apply the theorem from Appendix
\ref{DodatakMatricneTeoreme}, which states the following:
\begin{itemize}
\item $R=r+\rho$, where $R$, $r$ and $\rho$ are ranks of matrices $\Delta$,
$S''$ and $\tS$ respectively,
\item if $\rho = J$ then $\det M_{\Delta} = \pm \det M \det 2\tS $,
\item if $0<\rho<J$ then $\det M_{\Delta} (\det B_4)^2 = \pm \det M \det
M_{2\tS}$,
\end{itemize}
where the signs depend on even/odd rank of $\tS$. See Appendix
\ref{DodatakMatricneTeoreme} for a proof of the theorem and the definition of
the matrix $B_4$.

The $\Delta$ matrix can be expanded into power series
$$
\Delta = \Delta_0 + \frac{1}{\Lambda}\Delta_1 + \frac{1}{\Lambda^2} \Delta_2 +
\dots\; ,
$$
and as we shall see in the next section, the first nonzero leading term in the
series is always $\Delta_0$. Consequently the
determinant for the $\Delta$ matrix is given by
$$
\det\Delta = \det\Delta_0 + O \left( \frac{1}{\Lambda} \right) \,.
$$
Given that $M$ is a submatrix of $\Delta$, it follows that
$$
\det M = \det M_0 + O \left( \frac{1}{\Lambda} \right).
$$
Assume now that
$$
\tS = \frac{1}{\Lambda^n} \tS_n +  O \left( \frac{1}{\Lambda^{n+1}} \right).
$$
where $n\in \prirodni_0$. If the rank $\rho$ of $\tS$ is positive, we have
\begin{equation} \label{DeterminantaZaStildaMatricu}
\det (2\tS) = \frac{1}{\Lambda^{n\rho}} \det (2\tS_n) + O \left(
\frac{1}{\Lambda^{n\rho+1}} \right)\,.
\end{equation}

Equation (\ref{DeterminantaZaStildaMatricu}) implies three distinct
possibilities.
If $\rho=J$, then we can use the first identity for determinants from the
theorem, and obtain the equation
$$
\det\Delta_0 = \pm \frac{1}{\Lambda^{nJ}} \det M \det (2\tS_n) + O
\left( \frac{1}{\Lambda} \right),
$$
which is consistent if and only if $n=0$. If $0<\rho<J$,  we can use the second
identity for determinants, and obtain the equation
$$
\det\Delta_0 (\det B_4)^2 = \pm \frac{1}{\Lambda^{n\rho}} \det M
\det (2\tS_n) + O \left( \frac{1}{\Lambda} \right)\,,
$$
which is consistent if and only if $n=0$, due to the fact that $\det B_4 \sim
O(1)$ (see remark 3 in Appendix \ref{DodatakMatricneTeoreme}). Finally, if
$\rho=0$ we have $\tS=0$.

Therefore  we have essentially two possible situations. If $\rho=0$, then the
matrix $\tS$ is equal to zero, because $\rho$ is its rank. This implies that
the wavefunction (\ref{IntegraljenaTalasnaFunkcijaPrekoEksponenata}) is constant
in the leading order of $\Lambda$, i.e. the $j$ dependence only appears in the
subleading terms, which are of the type $(j-j_0)^3$ and higher. On the other
hand, if $n=0$, $\tS$ is different from zero in the leading order, so that
(\ref{IntegraljenaTalasnaFunkcijaPrekoEksponenata}) is a Gaussian, but not of
the required type (\ref{gas}). Hence the large-spin asymptotics of $\Psi_+$ is
never a Gaussian function of type (\ref{gas}).

\section{\label{SecComputation}Computation of the matrix $\tS$}

The main result of the previous section has been obtained under the assumption
that the matrix $\Delta$
has a leading contribution of $O(1)$ in $\Lambda$. In this section we will
demonstrate this by an explicit computation for some concrete spin-network
diagrams. In order to do so, we need to
explicitly find an extremal point $y_N^*$ and evaluate the corresponding
$\Delta$ for a given spin network.

An extremal point is defined by
 \begin{equation} \label{UsloviDaEkstremumBudeMaksimum}
\frac{\del S}{\del y_N} =0, \qquad \text{ where } y^* \in \Delta' , \qquad
N\in\cN \,,
 \end{equation}
where we must also take into account the consistency conditions
(\ref{JnaZaStacionarneTackeA}) and (\ref{JnaZaStacionarneTackeB}).

Differentiating (\ref{RazvojMatriceSpoLambda}) and (\ref{KomponenteMatriceS}) we
obtain
 \begin{equation} \label{OpstaJnaZaMinimum}
\frac{\del S_0}{\del y_N} + \frac{1}{\Lambda} \frac{\del S_1}{\del y_N} +
\frac{1}{\Lambda^2} \frac{\del S_2}{\del y_N} =
\cO\left(\frac{1}{\Lambda^3}\right),
 \end{equation}
$$
\begin{array}{ccl}
\ds \frac{\del S_0}{\del y_N} & = & \ds -2C (y_N-y_0) \delta_{N,\lambda} -
\sum_{\substack{v \\ N\in v}} \theta_{N,v} \left( \sum_{s\in v} y_s \theta_{s,v}
- m_v\pi \right) , \\
\ds \frac{\del S_1}{\del y_N} & = & \ds - \sum_{\substack{v \\ N\in v}}
\theta_{N,v} \left( \frac{\pi}{4} - n_v\pi \right), \\
\ds \frac{\del S_2}{\del y_N} & = & \ds \frac{1}{y_N} \delta_{N,a} -
\sum_{\substack{v \\ N\in v}} \left[ \frac{1}{2V_v}\frac{\del V_v}{\del y_N} +
\theta_{N,v} \DL + \left( \sum_{s\in v} y_s \theta_{s,v} -m_v\pi \right)
\frac{\del \DL}{\del y_N} \right]\,, \\
\end{array}
$$
where we have used the Schl\" afli differential identity
(\ref{SchlafliIdentitet}) for a tetrahedron
$$
\sum_{s\in v} y_s \frac{\del \theta_{s,v}}{\del y_N} = 0, \qquad \forall N,v \,.
$$

We will also need second derivatives:
 \begin{equation} \label{DrugiIzvodSpoY}
\begin{array}{ccl}
\ds \frac{\del^2 S_0}{\del y_M \del y_N} & = & \ds -2C \delta_{MN}
\delta_{N,\lambda} - \sum_{\substack{v \\ M,N\in v}} \left[
\theta_{M,v}\theta_{N,v} + \frac{\del\theta_{N,v}}{\del y_M} \left( \sum_{s\in
v} y_s \theta_{s,v} - m_v\pi \right) \right] , \\
\ds \frac{\del^2 S_1}{\del y_M \del y_N} & = & \ds - \sum_{\substack{v \\ M,N\in
v}} \frac{\del \theta_{N,v}}{\del y_M} \left( \frac{\pi}{4} - n_v\pi \right), \\
\ds \frac{\del^2 S_2}{\del y_M \del y_N} & = & \ds -\frac{1}{y_N^2} \delta_{MN}
\delta_{N,a} - \sum_{\substack{v \\ M,N\in v}} \left[ \frac{1}{2V_v}\frac{\del^2
V_v}{\del y_M\del y_N}
-\frac{1}{2V_v^2}\frac{\del V_v}{\del y_M}\frac{\del V_v}{\del y_N} + \frac{\del
\theta_{N,v}}{\del y_M} \DL \vphantom{\left( \sum_{s\in v} \right)} + \right. \\
 & & \ds \left. + \theta_{N,v} \frac{\del \DL}{\del y_M} + \theta_{M,v}
\frac{\del \DL}{\del y_N}
+ \left( \sum_{s\in v} y_s \theta_{s,v} -m_v\pi \right) \frac{\del^2 \DL}{\del
y_M\del y_N} \right] . \\
\end{array}
 \end{equation}

A solution of the system (\ref{OpstaJnaZaMinimum}) can be obtained
perturbatively in $1/\Lambda$ via the ansatz (\ref{FundamentalniAnsatz}),
$$
y_N^* = A_N + \frac{B_N}{\Lambda} + \frac{C_N}{\Lambda^2} + O\left(
\frac{1}{\Lambda^3} \right)\,.
$$

In the lowest order we obtain a nonlinear system for $A_N$,
$$
2C(A_N-y_0) \delta_{N,\lambda} + \sum_{\substack{v \\ N\in v}} \theta_{N,v}(A)
\left( \sum_{s\in v} A_s \theta_{s,v}(A) - m_v \pi \right) =0,
$$
which can be reduced to a simple equation using (\ref{JnaZaStacionarneTackeA}),
 \begin{equation} \label{JnaZaAove}
A_{\lambda} = y_0 .
 \end{equation}
This was already guessed before based on the analysis that the extremal point
must be in the vicinity of extremal points of the cosine and the Gaussian
functions $\mu(x_{\lambda})$.

At $O(1/\Lambda)$ order we obtain also a linear system of equations for
the $B_N$ coefficients
$$
\sum_M \left[ 2C \delta_{MN} \delta_{N,\lambda}
+ \sum_{\substack{v \\ M,N\in v}} \frac{\del \theta_{N,v}}{\del y_M}\Big|_{A}
\left( \sum_{s\in v} A_s \theta_{s,v}(A) - m_v \pi \right)
+ \sum_{\substack{v \\ M,N\in v}} \theta_{N,v}(A) \theta_{M,v}(A) \right] B_M =
$$
$$
\hphantom{mmmmmmmmmmmmmmmmmmmmmmmm}
= - \sum_{\substack{v \\ N\in v}} \theta_{N,v}(A) \left( \frac{\pi}{4} - n_v\pi
\right),
$$
which can be simplified using (\ref{JnaZaStacionarneTackeA}) and
(\ref{JnaZaStacionarneTackeB}) to
 \begin{equation} \label{JnaZaBove}
B_{\lambda} = 0.
 \end{equation}

Similarly, by applying (\ref{JnaZaStacionarneTackeA}) and
(\ref{JnaZaStacionarneTackeB}) at the $O(1/\Lambda^2)$ order we obtain a
linear system for the coefficients $C_N$
 \begin{equation} \label{JnaZaCove}
\sum_M \left[ 2C \delta_{MN} \delta_{N,\lambda} + \sum_{\substack{v \\ M,N\in
v}} \theta_{N,v}(A) \theta_{M,v}(A) \right] C_M =
\hphantom{mmmmmmmmmmmmm}
 \end{equation}
$$
\hphantom{mmmmmmmmmmmmm}
= -\frac{1}{A_N} \delta_{N,a} - \sum_{\substack{v \\ N\in v}} \left(
\frac{1}{2V_v}\frac{\del V_v}{\del y_N}\Big|_A + \theta_{N,v}(A) \DL(A) \right).
$$

Let us now write the complete set of equations which determine
the extremal point $y^*$
\begin{itemize}
\item equations for $A_N$:
 \begin{equation} \label{SistemJnaZaStacionarneTackeA}
\sum_{s\in v} A_s \theta_{s,v}(A) = m_v \pi, \qquad A_{\lambda} = y_0,
 \end{equation}
\item equations for $B_N$:
 \begin{equation} \label{SistemJnaZaStacionarneTackeB}
\sum_{s\in v } B_s \theta_{s,v}(A) = n_v\pi - \frac{\pi}{4}, \qquad B_{\lambda}
= 0,
 \end{equation}
\item equations for $C_N$:
 \begin{equation} \label{SistemJnaZaStacionarneTackeC}
\sum_M \left[ 2C \delta_{MN} \delta_{N,\lambda}
+ \sum_{\substack{v \\ M,N\in v}} \theta_{N,v}(A) \theta_{M,v}(A) \right] C_M =
\hphantom{mmmmmmmmmmmmm}
 \end{equation}
$$
\hphantom{mmmmmmmmmmmmm}
-\frac{1}{A_N} \delta_{N,a}
- \sum_{\substack{v \\ N\in v}} \left( \frac{1}{2V_v}\frac{\del V_v}{\del
y_N}\Big|_A + \theta_{N,v}(A) \DL(A) \right)
$$
\end{itemize}

Before we engage in finding solutions to these equations, let us introduce
some notation and discuss the form of the second derivatives of the phase
$S(\Lambda, y)$. We will introduce the following shorter notation for various
derivatives evaluated at the particular extremal point:
$$
\theta_{N,v} = \theta_{N,v}(A), \qquad \theta_{MN,v} =
\frac{\del\theta_{N,v}}{\del y_M} \Big|_{A}, \qquad
\theta_{MNs,v} = \frac{\del^2\theta_{N,v}}{\del y_s\del y_M} \Big|_A, \qquad
\theta_{MNpq,v} = \frac{\del^3\theta_{N,v}}{\del y_p \del y_q \del y_M} \Big|_A,
$$
$$
V_v = V_v (A), \qquad V_{N,v} = \frac{\del V_v}{\del y_N} \Big|_A, \qquad
V_{MN,v} = \frac{\del^2 V_v}{\del y_M\del y_N} \Big|_A.
$$

By using (\ref{DrugiIzvodSpoY}) and the ansatz (\ref{FundamentalniAnsatz}) for
$y^*$,
as well as the equations (\ref{JnaZaStacionarneTackeA}) and
(\ref{JnaZaStacionarneTackeB}), we obtain
$$
\Delta_{MN} = - \left\{ 2C \delta_{MN} \delta_{N,\lambda} + \sum_{\substack{v \\
M,N\in v}} \theta_{M,v}\theta_{N,v} \right\}
-\frac{1}{\Lambda} \left\{ \sum_{\substack{v \\ M,N\in v}} \sum_{s\in v} B_s
\left( \theta_{M,v} \theta_{Ns,v} + \theta_{N,v} \theta_{Ms,v} \right)
\right\} -
$$
$$
- \frac{1}{\Lambda^2} \left\{
\frac{1}{A_N^2} \delta_{MN} \delta_{N,a} + \sum_{\substack{v \\ M,N\in v}}
\left[ \frac{V_{MN,v}}{2V_v} -\frac{V_{M,v}V_{N,v}}{2V_v^2} + \theta_{MN,v} \DL
+ \theta_{N,v} \frac{\del \DL}{\del y_M}\Big|_A + \theta_{M,v} \frac{\del
\DL}{\del y_N}\Big|_A  \right]
\right.
$$
$$
+\!\! \sum_{\substack{v \\ M,N\in v}} \! \left[ \!
\theta_{M,v} \sum_{s\in v} C_s \theta_{Ns,v} + \theta_{N,v} \sum_{s\in v} C_s
\theta_{Ms,v} + \frac{1}{2} \theta_{M,v} \sum_{p,q\in v} B_p B_q \theta_{Npq,v}
+ \frac{1}{2} \theta_{N,v} \sum_{p,q\in v} B_p B_q \theta_{Mpq,v} \! \right]
$$
 \begin{equation} \label{DeltaMatrica}
+ \left. \sum_{\substack{v \\ M,N\in v}} \theta_{MN,v} \left[ \sum_{s\in v} C_s
\theta_{s,v} + \left( \frac{\pi}{4} - n_v\pi \right) \sum_{s\in v}
B_s\theta_{MNs,v} + \frac{1}{2} \sum_{p,q\in v} B_p B_q \theta_{pq,v} \right]
\right\} + \cO\left( \frac{1}{\Lambda^3} \right).
\end{equation}

Therefore, after finding an explicit extremal point, we have substituted the
coefficients $A_N$, $B_N$ and $C_N$ into the above equations and obtained an
explicit expression for the $\Delta$ matrix. The curly braces group all terms of
orders $O(1)$, $O(1/\Lambda)$ and $O(1/\Lambda^2)$, respectively. This
demonstrates
that the leading order of the $\Delta$ matrix is an
$O(1)$ term.

Given that all of the above equations are fairly complicated, we will
investigate
them order by order in $\Lambda$.

\subsection{\label{SubsekcijaVodeciRed}The $O(1)$ approximation}

Let us rewrite the equation (\ref{SistemJnaZaStacionarneTackeA}) in this
approximation as
 \begin{equation} \label{SistemJnaZaStacionarneTackeAuVodecaAprox}
\sum_{s\in v} A_s \theta_{s,v}(A) = m_v \pi , \qquad A_{\lambda} = y_0 \,.
 \end{equation}
We will also rewrite the expression for the $\Delta$ matrix as
 \begin{equation} \label{DeltaMatricaVodecaAprox}
\Delta_{MN} = -  2C \delta_{MN} \delta_{N,\lambda} - \sum_{\substack{v \\ M,N\in
v}} \theta_{M,v}(A)\theta_{N,v}(A) .
 \end{equation}

The calculation of the matrix $\tilde S$ can be organized in the following way:
\begin{itemize}
\item calculate all extremal points $y_N^*=A_N$ by solving equation
(\ref{SistemJnaZaStacionarneTackeAuVodecaAprox});
\item for every extremal point obtained, calculate $\Delta$ matrix according to
(\ref{DeltaMatricaVodecaAprox});
\item split the $\Delta$ matrix to blocks $\ddot{S}$, $\dot{S}'$ and $S''$;
\item determine the rank of $S''$, and the null-space projector $K$ if the rank
is less than $\omega$;
\item check whether or not $\dot{S}' K =0$; if it is nonzero the procedure fails
and the wavefunction does not have Gaussian form;
\item determine an orthogonal matrix $O$ which diagonalizes $K$ and use it to
change to a basis where matrices $S''$ and $\dot{S}'$ are block-diagonal;
read-off the nonzero blocks $M$ and $N$;
\item compute the matrix $\tilde{S}$ according to the equation
$$
\tS \equiv - \frac{1}{2} \left( \ddot{S} - N M^{-1}N^T \right).
$$
\end{itemize}

The system of equations (\ref{SistemJnaZaStacionarneTackeAuVodecaAprox}) is
highly nonlinear, and therefore extremely hard to solve. Nevertheless, one exact
solution can be guessed
$$
y_N^* = A_N = y_0, \qquad \theta_{N,v} = \theta_0  \equiv \pi - \arcsin
\frac{2\sqrt{2}}{3} \equiv \arccos \left( -\frac{1}{3} \right), \qquad m_v =
\frac{6\theta_0}{\pi} y_0.
$$

We will refer to this solution as the ``diagonal'' solution. It corresponds to a
situation where all the tetrahedra are equilateral, which is a very symmetrical
configuration.
This greatly simplifies the equations, and the required matrices can be
calculated without additional approximations. In the case of the loop spin
network and the theta spin network a computer calculation gives
 \begin{equation}
\dot{S}' K = 0, \qquad R=r, \qquad \tS = 0 \,.\label{rzero}
 \end{equation}
This result agrees with the $\rho=0$ case discussed in section
\ref{SecAsymptotics}.

The calculation of the $\tilde S$ matrix can be also performed numerically for
non-diagonal solutions. In
fact, in the numerical approach, the hardest first step, solving the equation
(\ref{SistemJnaZaStacionarneTackeAuVodecaAprox}), can be completely
sidestepped. This is due to the fact that $\Delta$ depends on $A_N$ only through
the angles $\theta_{N,v}(A)$. Thus one can design an algorithm which chooses the
angles completely randomly from their domain $[0,\pi]$, which in principle
covers also the angles obtained using any specific solution of
(\ref{SistemJnaZaStacionarneTackeAuVodecaAprox}). The constant $C>0$ can be
also chosen randomly.

One such algorithm has been implemented on a computer to explicitly calculate
the $\Delta$ matrix, the projector $K$, the rank of $S''$, then $\dot{S}'K$, $M$
and $N$, and finally $\tS$. It has been executed $50$ times with random initial
data for the cases of a loop spin network and a theta spin network, and each
execution gave the same result (\ref{rzero}).
The numerical precision of the calculation was $10^{-6}$. The precision
can be arbitrarily increased at the expense of the execution time, and in $5$
executions the precision was raised to $10^{-15}$, with no change in the result.

\subsection{\label{SubsekcijaPrviPopravljeniRed}The $O(1/\Lambda)$
approximation}

Let us rewrite the all necessary equations up to $O(1/\Lambda^2)$. The extremal
points can be found using the ansatz
$$
y_N^* = A_N + \frac{B_N}{\Lambda}\,.
$$
The coefficients $A_N$ and $B_N$ will be determined by the
equations
 \begin{equation} \label{SistemJnaZaStacionarneTackeAPrvaPopr}
\sum_{s\in v} A_s \theta_{s,v}(A) = m_v \pi, \qquad A_{\lambda} = y_0\,,
 \end{equation}
 \begin{equation} \label{SistemJnaZaStacionarneTackeBPrvaPopr}
\sum_{s\in v } B_s \theta_{s,v}(A) = n_v\pi - \frac{\pi}{4}, \qquad B_{\lambda}
= 0\,.
 \end{equation}
$\Delta$ matrix is given as:
 \begin{equation} \label{DeltaMatricaPrvaPopr}
\Delta_{MN} = - 2C \delta_{MN} \delta_{N,\lambda} - \sum_{\substack{v \\ M,N\in
v}} \theta_{M,v}\theta_{N,v}
-\frac{1}{\Lambda} \left\{ \sum_{\substack{v \\ M,N\in v}} \sum_{s\in v} B_s
\left( \theta_{M,v} \theta_{Ns,v} + \theta_{N,v} \theta_{Ms,v} \right)
\right\}\,,
 \end{equation}
where
$$
\theta_{N,v} = \theta_{N,v}(A), \qquad \theta_{Ns,v} = \frac{\del
\theta_{N,v}}{\del y_s} \Big|_A \,.
$$

The procedure for computation of $\tilde{S}$ matrix is the same as in the
$O(1)$ approximation, up to two additional steps. These two steps consist of
solving the linear system of
equations (\ref{SistemJnaZaStacionarneTackeBPrvaPopr}) for $B_N$ coefficients,
and then expanding the resulting matrix into a power series in $1/\Lambda$,
$$
\tS = \tS_0 + \frac{\tS_1}{\Lambda} + O \left(\frac{1}{\Lambda^2}\right) \,.
$$

This case can be also analyzed analytically (by using the diagonal solution), as
well as numerically (by randomly generating the $A_N$ coefficients). In contrast
to the
previous case, one should also always specify the $n_v$ parameters as a part of
the initial
data, in order to solve (\ref{SistemJnaZaStacionarneTackeBPrvaPopr}). One
algorithm for this has also been implemented, and we obtained the same result
(\ref{rzero}) for a loop and a theta  spin network.

\section{\label{SecPsiMinus}$\Psi_-$ integral}

The asymptotic analysis of $\Psi_-$ can be done by using the same method as
in the $\Psi_+$ case. However, it turns out that $\Psi_-$ analysis is
considerably simpler due to qualitatively different nature of the scaling laws
for the asymptotic expressions for $6j$ symbols when some of the spins are not
large. 

The $\Psi_-$ part of the wavefunction can be approximated by an analogous expression to (\ref{lambint}) 
\begin{equation}
\Psi_- (j,j_0)\approx 2^A \Lambda^{A+\omega} \sum_{z} F(z) \int_{D_-'}
d^{\omega}y \; \prod_a y_a \prod_\alpha y_\alpha \prod_{\lambda}
\frac{e^{-\Lambda^2 (y_{\lambda}-y_0)^2}}{y_\lambda}
\prod_v \{ 6j_v(\Lambda y,z) \}\,,
\label{psiminus} \end{equation}
where the vector $z$ denotes the small spins, while the vector $\Lambda y$
denotes the large ones. We will sum over the small spins rather than integrate
and the
number $\omega$ of large spins is smaller than in the $\Psi_+$ case, while $F(z)$ represents
the
part of the wavefunction which does not depend on large spins $\Lambda y$. Which
spins can remain small and
which must be large depends on the triangle inequalities built into
the
$6j$ symbols. These restrictions on the spins $z$ and $\Lambda y$ will also depend on the detailed topology of the spin
network. However, these details will not affect the asymptotic analysis.

The next step is to use the asymptotic formulas for the $6j$ symbols, but now
we cannot use only the PR formula (\ref{AsimptotikaSestJ}) since the $6j$ symbols where not all of the spins are large will appear. The asymptotics of $6j$ symbols with $3$, $4$ or $5$ large 
spins is given in the Appendix, see formulas
(\ref{AsimptotikaPetJ}), (\ref{AsimptotikaCetiriJ}) and
(\ref{AsimptotikaTriJ}).

Let the asymptotic behavior of each $6j$ symbol be described by a function
$\phi_i (\Lambda y,z)$, so that
$$
\{ 6j(\Lambda y,z) \} \approx \phi_i (\Lambda y,z) \,\,
\textrm{as}\,\,\Lambda\to\infty\,,
$$
where $i=3,4,5,6$ denotes the number of large spins. When $i<6$, we will refere to the corresponding $6j$ symbols as degenerate. When $i=6$, we will refere to the corresponding  $6j$ symbol as a non-degenerate. 

Note that in a degenerate case
\begin{equation}\phi_i (\Lambda y,z) \approx \Lambda^{\rho_i}\,f_i (y,z) \,\,
\textrm{as}\,\,\Lambda\to\infty\,,\label{smspa}\end{equation}
while in the non-degenarte case
\begin{equation}\phi (\Lambda y) \approx \Lambda^{\rho} \cos \left( \Lambda
f(y)\right) \,\, \textrm{as}\,\,\Lambda\to\infty\,.\label{lspa}\end{equation}

Let us write
the integrand of $\Psi_-$ integral in an exponential form suitable for stationary
point approximation. The corresponding $\Delta$ matrix is given as before
$$
\Delta^{-}_{MN} = \frac{\del^2 S_-(\Lambda y,z)}{\del y_M \del y_N}\Big|_{y=y^*}
\,,
$$
where the new phase $S_-$ is now proportional to
$\sum_v \ln \phi_v(\Lambda y,z)$. It is easy to see that the $\Delta^-$
matrix will contain the terms of the form
$$
\left( \frac{1}{\phi_v} \frac{\del \phi_v}{\del \Lambda y} \right)^2
$$
and 
$$
\frac{1}{\phi_v} \frac{\del^2 \phi_v}{\del \Lambda y \del \Lambda y} \,.
$$

All these terms are of $O(1/\Lambda^2)$, since the asymptotic
functions $\phi_v$ behave well when differentiated by $\Lambda y$, see
(\ref{smspa}). 
Furthermore, given that the $\Delta$ matrix should be evaluated at an extremal
point $y^*$, one can see that the $6j$ symbols which contain only $3$ or $4$
large
spins give a sub-leading contribution, since they decrease monotonically
in the limit $\Lambda \to \infty $ and thus do not have any extremal points. The
$6j$ symbols with $5$ large spins have extremal points, but they also provide
only a contribution of $O(1/\Lambda^2)$ to the $\tilde{S}_-$ matrix. The $O(1)$ contribution to 
$\tilde{S}_-$ comes only from
those $6j$ symbols where all spins are large, due to the fact that the argument
of the cosine function in (\ref{lspa}) is proportional to $\Lambda$.

Therefore the $\Psi_-$ integral will have the same asymptotics as the
$\Psi_+$ integral if there is at least one non-degenerate $6j$ symbol contributing to $\tilde{S}_-$. If only the degenerate $6j$ symbols are present, the contribution to $\tilde{S}_-$ will be of $O(1/\Lambda^2)$. 

\section{\label{SecConclusion}Conclusions}

We have shown that the leading term in the large-spin asymptotics of a
flat-space wavefunction $\Psi_k (\Gamma, j_0)\equiv\Psi (j,j_0)$ is given by
 \begin{equation} \Psi (j,j_0) \approx \sum_{n,p} N_{np}^- (j_0)
\,e^{-\frac{1}{2}(j - j_0)^T (B^-_{np} + \frac{1}{j_0^2} D^-_{np})(j - j_0 )} +
\sum_{n,q} N_{nq}^+ (j_0) \,e^{-\frac{1}{2}(j - j_0)^T B^+_{nq} (j - j_0 )}
\,,\label{fina} \end{equation}
where $n\in \bf{Z}$, $p\in S_-$, $q\in S_+$ and  $B$ and $D$ are constant
(independent of $j_0$) matrices. The index sets $S_\pm$ correspond to the
stationary points of the large-spin approximations $f_\pm$ of $f$ in the regions
$D_\pm$ and the corresponding functions $N^\pm (j_0)$ will be powers
$j_0^{r_\pm}$, $r_\pm \in \bf Q$. In the case of a loop and a theta spin network
the computer results are consistent with $B_{nq}^+ = C_{nq}^+ = 0$ for all $n$ and all $q$ which means that in those cases the matrix $\tilde S$ vanishes.

The asymptotics (\ref{fina}) is not of the type (\ref{gas}) required
for the correct semi-classical limit. This means that the wavefunction $\Psi_k
(\Gamma,j_0)$ will not give the correct graviton propagator asymptotics. One can
argue that some other wavefunction may give the correct asymptotics, but the
problem is to see what other wavefunction can replace $\Psi_k (\Gamma,j_0)$.
Although our result applies only to the Euclidean LQG, one wonders what is the
relevance of this result for the Lorentzian LQG, given that there is a strong
belief that the Euclidean and the Lorentzian theories should be related by some
kind of an analytic continuation. Note that Lorentzian analogs of the Euclidean
wavefunctions used in this paper are not known. However, one can try use one of
the recently proposed Lorentzian spin foam models \cite{epr,elpr,fk} in order to
construct a Lorentzian spin-network  wavefunction. The large-spin asymptotics
could be then computed by using essentially the same techniques as the ones
introduced in this paper.

As far as the our result is concerned, there are certain caveats. The first
caveat is that the obtained asymptotics is for the Ponzano-Regge regularization
$\Psi_k^{PR}(\Gamma,j_0)$ of the zero-cosmological constant spin-network
wavefunction $\Psi (\Gamma, j_0)$. The wavefunction $\Psi_k^{PR}(\Gamma,j_0)$
is different from the quantum group regularization $\Psi_k (\Gamma,j_0)$ of
$\Psi(\Gamma,j_0)$, but the physics intuition suggests that the asymptotic
behavior of $\Psi_k$ and $\Psi_k^{PR}$ for large $k$ should be essentially the
same, up to constant factors, see for example \cite{fl}. However, it still
remains to be proven that these two wavefunctions have the same semiclassical
asymptotics.

The second caveat is that the construction of $\Psi_k$ wavefunctions is
triangulation dependent and our calculations have been done for the simplest
triangulation of $S^3$. However, it is not difficult to see that by taking a
more complex triangulation, the corresponding asymptotics will not change 
qualitatively, because the corresponding $f(x,j,j_0)$ will be always a product
of $6j$ symbols and our method of computing the asymptotics is independent of
the number of $6j$ symbols. Similarly, one can hope that a special choice of the function
$\mu(\lambda)$ can lead to a desired asymptotics. However, a generic $\mu(\lambda)$ gives a 
contribution to $\tilde S$ of $O(1)$ in $j_0$. Note that it is possible to fine-tune $\mu$ for a given spin network such
that the $O(1)$ contributions cancel and the $O(1/j_0)$ contributions are non-zero. However, this fine-tunning depends on the spin network and hence one cannot find a $\mu$ which will work for all spin networks.

\bigskip
\bigskip
\noindent{\bf\Large Acknowledgments}

\bigskip
We would like to thank John W. Barrett for discussions. AM was partially
supported by the FCT grants PTDC/MAT/69635/2006 and PTDC/MAT/099880/2008. MV was
supported by the FCT grants SFRH/BPD/46376/2008 and PTDC/MAT/099880/2008.

\bigskip
\bigskip
\noindent{\bf\Large Appendix}

\appendix

\section{The spin-network diagram}

The spin-network whose evaluation appears in (\ref{sixjr}), is given in the case
of a single loop spin network with spin $j$ by the following diagram

\begin{center}
\setlength{\unitlength}{1.5mm}
\begin{picture}(78,90)
\thinlines
\drawpath{22.0}{80.0}{54.0}{80.0}
\Thicklines
\drawpath{22.0}{54.0}{54.0}{54.0}
\thinlines
\drawpath{22.0}{74.0}{54.0}{60.0}
\drawpath{22.0}{48.0}{54.0}{48.0}
\drawpath{54.0}{74.0}{40.0}{68.0}
\drawpath{22.0}{60.0}{36.0}{66.0}
\drawpath{22.0}{80.0}{22.0}{48.0}
\drawpath{54.0}{80.0}{54.0}{48.0}
\drawpath{40.0}{48.0}{40.0}{48.0}
\drawdotline{22.0}{48.0}{22.0}{42.0}
\drawdotline{54.0}{48.0}{54.0}{42.0}
\drawpath{22.0}{42.0}{54.0}{42.0}
\Thicklines
\drawpath{22.0}{36.0}{54.0}{36.0}
\thinlines
\drawpath{54.0}{30.0}{22.0}{16.0}
\drawpath{54.0}{16.0}{40.0}{22.0}
\drawpath{22.0}{30.0}{36.0}{24.0}
\drawpath{22.0}{10.0}{54.0}{10.0}
\drawpath{22.0}{42.0}{22.0}{10.0}
\drawpath{54.0}{42.0}{54.0}{10.0}
\drawdotline{22.0}{80.0}{14.0}{60.0}
\drawdotline{14.0}{60.0}{14.0}{30.0}
\drawdotline{14.0}{30.0}{22.0}{10.0}
\drawdotline{54.0}{80.0}{62.0}{60.0}
\drawdotline{62.0}{60.0}{62.0}{30.0}
\drawdotline{62.0}{30.0}{54.0}{10.0}
\drawpath{8.0}{86.0}{8.0}{4.0}
\drawpath{8.0}{4.0}{68.0}{4.0}
\drawpath{68.0}{4.0}{68.0}{86.0}
\drawpath{68.0}{86.0}{8.0}{86.0}
\drawlefttext{14.0}{46.0}{$\lambda_1$}
\drawrighttext{22.0}{44.0}{$\lambda_2$}
\drawlefttext{54.0}{44.0}{$\lambda_3$}
\drawrighttext{62.0}{46.0}{$\lambda_4$}
\drawcenteredtext{38.0}{56.0}{$j$}
\drawcenteredtext{38.0}{34.0}{$j$}
\drawcenteredtext{38.0}{50.0}{$a$}
\drawcenteredtext{38.0}{40.0}{$a$}
\drawlefttext{54.0}{66.0}{$b$}
\drawlefttext{54.0}{24.0}{$b$}
\drawcenteredtext{38.0}{82.0}{$c$}
\drawcenteredtext{38.0}{8.0}{$c$}
\drawrighttext{22.0}{66.0}{$d$}
\drawrighttext{22.0}{24.0}{$d$}
\drawcenteredtext{30.0}{72.0}{$e$}
\drawcenteredtext{30.0}{18.0}{$e$}
\drawcenteredtext{46.0}{72.0}{$f$}
\drawcenteredtext{46.0}{18.0}{$f$}
\drawcenteredtext{12.0}{82.0}{$\alpha$}
\drawcenteredtext{16.0}{56.0}{$\beta$}
\drawcenteredtext{60.0}{56.0}{$\gamma$}
\drawcenteredtext{38.0}{76.0}{$\delta$}
\drawcenteredtext{28.0}{66.0}{$\varepsilon$}
\drawcenteredtext{48.0}{66.0}{$\mu$}
\drawcenteredtext{38.0}{60.0}{$\nu$}
\drawcenteredtext{44.0}{52.0}{$\rho$}
\drawcenteredtext{44.0}{46.0}{$\sigma$}
\drawcenteredtext{44.0}{38.0}{$\tau$}
\drawcenteredtext{38.0}{30.0}{$\varphi$}
\drawcenteredtext{28.0}{24.0}{$\chi$}
\drawcenteredtext{48.0}{24.0}{$\xi$}
\drawcenteredtext{38.0}{14.0}{$\zeta$}
\drawlefttext{22.0}{78.0}{$\iota_1$}
\drawlefttext{22.0}{12.0}{$\iota_1$}
\drawrighttext{54.0}{78.0}{$\iota_2$}
\drawrighttext{54.0}{12.0}{$\iota_2$}
\drawlefttext{22.0}{58.0}{$\iota_3$}
\drawlefttext{22.0}{32.0}{$\iota_3$}
\drawrighttext{54.0}{58.0}{$\iota_4$}
\drawrighttext{54.0}{32.0}{$\iota_4$}
\drawlefttext{22.0}{50.0}{$\iota_5$}
\drawlefttext{22.0}{40.0}{$\iota_5$}
\drawrighttext{54.0}{50.0}{$\iota_6$}
\drawrighttext{54.0}{40.0}{$\iota_6$}
\end{picture}
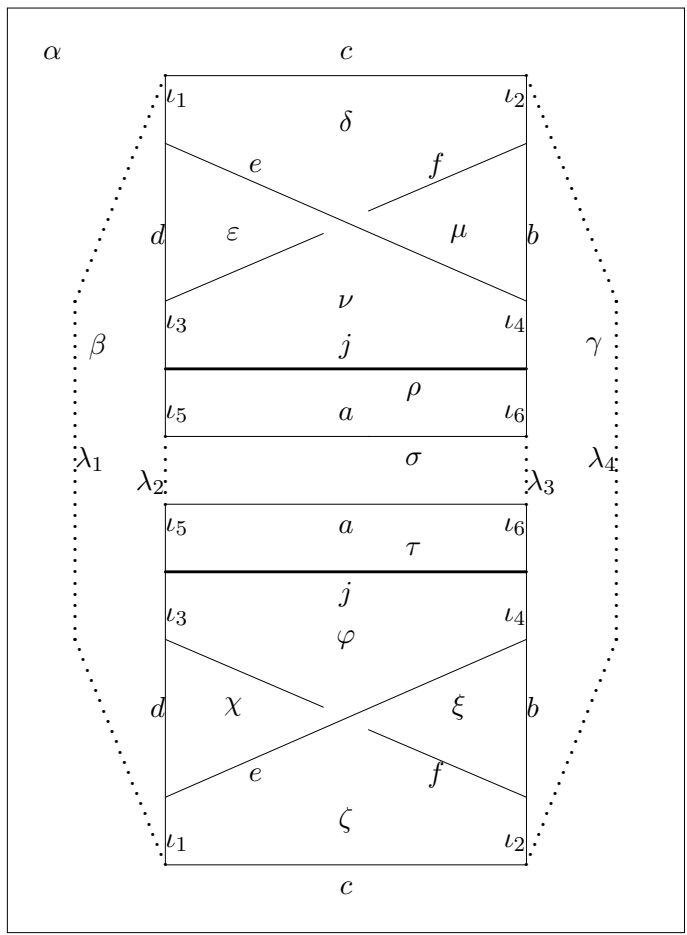
\\ Figure 3.
\end{center}
All spins take values from the set $\{ 0,\frac{1}{2},1,\frac{3}{2},2\dots,
\frac{k}{2} \}$. Here $n$ is the degree of the corresponding quantum group
$SU(2)_q$ such that $q^{2k+2}=1$ i.e. $q=e^{i\pi/k+2}$. 

The evaluation  of the above diagram, to which we refer as the amplitude, can be
calculated via the following rules:
\begin{itemize}
\item The amplitude of the whole diagram is the product of the amplitudes for
the
vertices.
\item The amplitude of a three-vertex is proportional to the amplitude of the
corresponding tetrahedron spin network, as follows
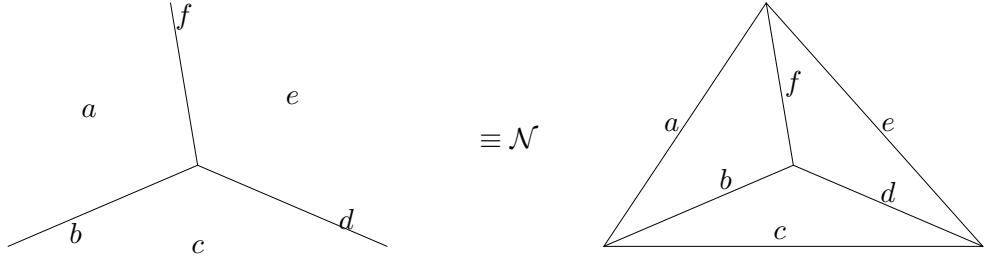
\begin{center}
\setlength{\unitlength}{0.9mm}
\begin{picture}(152,46)
\thinlines
\drawpath{116.0}{42.0}{92.0}{6.0}
\drawpath{92.0}{6.0}{148.0}{6.0}
\drawpath{148.0}{6.0}{116.0}{42.0}
\drawpath{116.0}{42.0}{120.0}{18.0}
\drawpath{120.0}{18.0}{92.0}{6.0}
\drawpath{120.0}{18.0}{148.0}{6.0}
\drawpath{32.0}{18.0}{28.0}{42.0}
\drawpath{32.0}{18.0}{4.0}{6.0}
\drawpath{32.0}{18.0}{60.0}{6.0}
\drawcenteredtext{16.0}{26.0}{$a$}
\drawcenteredtext{14.0}{8.0}{$b$}
\drawcenteredtext{32.0}{6.0}{$c$}
\drawcenteredtext{54.0}{10.0}{$d$}
\drawcenteredtext{46.0}{28.0}{$e$}
\drawcenteredtext{30.0}{40.0}{$f$}
\drawcenteredtext{102.0}{24.0}{$a$}
\drawcenteredtext{110.0}{16.0}{$b$}
\drawcenteredtext{118.0}{8.0}{$c$}
\drawcenteredtext{134.0}{14.0}{$d$}
\drawcenteredtext{134.0}{24.0}{$e$}
\drawcenteredtext{120.0}{30.0}{$f$}
\drawcenteredtext{78.0}{22.0}{$\equiv {\cal N}$}
\end{picture}
\end{center}
\item The amplitude for a four-vertex is proportional to the amplitude of the
corresponding tetrahedron spin network, which is given by
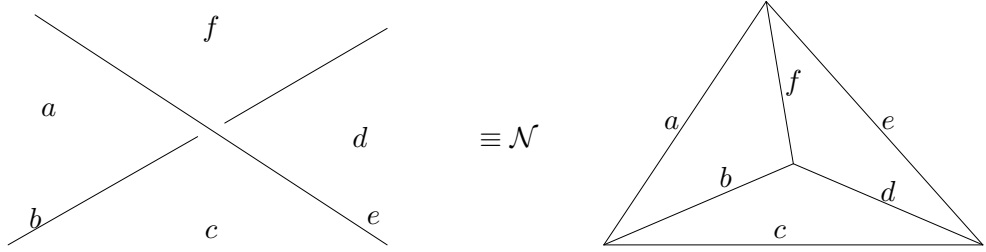
\begin{center}
\setlength{\unitlength}{0.9mm}
\begin{picture}(152,44)
\thinlines
\drawpath{116.0}{40.0}{92.0}{4.0}
\drawpath{92.0}{4.0}{148.0}{4.0}
\drawpath{148.0}{4.0}{116.0}{40.0}
\drawpath{116.0}{40.0}{120.0}{16.0}
\drawpath{120.0}{16.0}{92.0}{4.0}
\drawpath{120.0}{16.0}{148.0}{4.0}
\drawcenteredtext{10.0}{24.0}{$a$}
\drawpath{4.0}{4.0}{32.0}{20.0}
\drawpath{36.0}{22.0}{60.0}{36.0}
\drawpath{8.0}{38.0}{60.0}{4.0}
\drawcenteredtext{8.0}{8.0}{$b$}
\drawcenteredtext{34.0}{6.0}{$c$}
\drawcenteredtext{56.0}{20.0}{$d$}
\drawcenteredtext{58.0}{8.0}{$e$}
\drawcenteredtext{34.0}{36.0}{$f$}
\drawcenteredtext{102.0}{22.0}{$a$}
\drawcenteredtext{110.0}{14.0}{$b$}
\drawcenteredtext{118.0}{6.0}{$c$}
\drawcenteredtext{134.0}{12.0}{$d$}
\drawcenteredtext{134.0}{22.0}{$e$}
\drawcenteredtext{120.0}{28.0}{$f$}
\drawcenteredtext{78.0}{20.0}{$\equiv {\cal N}$}
\end{picture}
\end{center}
\end{itemize}
The normalization ${\cal N}$ is given by, see \cite{marm}
$$
{\cal N} = \frac{1}{\sqrt{|\Theta(a,b,c) \Theta(c,d,e) \Theta(a,e,f)
\Theta(b,d,f) |}}
$$
where $\Theta(a,b,c)$ is the evaluation of the $\theta$-graph
$$
\Theta(a,b,c) \equiv (-1)^{a+b+c} \frac{[a+b-c]_q! [a+c-b]_q! [b+c-a]_q!
[a+b+c+1]_q!}{[2a]_q! [2b]_q! [2c]_q!}.
$$

\section{\label{AppTetrahedron}Tetrahedron spin network and the $6j$ symbol}

The amplitude for a tetrahedron spin network is, roughly
speaking, the value of the corresponding $6j$ symbol. Following the conventions
of \cite{mik2} and \cite{Carter1995}, the amplitude is given by the equation
 \begin{equation} \label{KonvencijaZaSestJ}
\rule[-11mm]{0mm}{11mm}
\text{\setlength{\unitlength}{0.8mm}
\begin{picture}(64,30)(0,14)
\thinlines
\drawpath{28.0}{40.0}{4.0}{4.0}
\drawpath{4.0}{4.0}{60.0}{4.0}
\drawpath{60.0}{4.0}{28.0}{40.0}
\drawpath{28.0}{40.0}{32.0}{16.0}
\drawpath{32.0}{16.0}{4.0}{4.0}
\drawpath{32.0}{16.0}{60.0}{4.0}
\drawcenteredtext{14.0}{22.0}{$a$}
\drawcenteredtext{22.0}{14.0}{$b$}
\drawcenteredtext{30.0}{6.0}{$c$}
\drawcenteredtext{46.0}{12.0}{$d$}
\drawcenteredtext{46.0}{22.0}{$e$}
\drawcenteredtext{32.0}{28.0}{$f$}
\end{picture}}
\equiv \quad \sestj{a}{b}{c}{d}{e}{f}_q \sqrt{|\Theta(a,b,c) \Theta(c,d,e)
\Theta(a,e,f) \Theta(b,d,f) |}\,.
 \end{equation}

The tetrahedron associated to a $6j$ symbol has the following geometric
properties.
The length of an edge colored by a spin $j$, is given by
$$
l=j+\frac{1}{2}\,.
$$
The area of a face, with edge lengths $l_1, l_2, l_3$, is given by the Heron
formula
$$
A_{123} = \sqrt{s(s-l_1)(s-l_2)(s-l_3)}, \qquad \text{ where } \qquad s\equiv
\frac{1}{2} (l_1+l_2+l_3)\,.
$$
The volume of the tetrahedron is given by the Tartaglia
determinant
$$
V^2 = \frac{1}{288} \det \left[
\begin{array}{ccccc}
       0 & l^2_{12} & l^2_{13} & l^2_{14} & 1 \\
l^2_{12} &        0 & l^2_{23} & l^2_{24} & 1 \\
l^2_{13} & l^2_{23} &        0 & l^2_{34} & 1 \\
l^2_{14} & l^2_{24} & l^2_{34} &        0 & 1 \\
       1 &        1 &        1 &        1 & 0 \\
\end{array}
\right],
$$
where $l_{ij}$ is the edge length between the vertices $i$ and $j$. Note that if
we
fix all $6$ edge lengths, we could create in total $6!$ different tetrahedra.
However,
some of them are equivalent up to rotations and reflections. When this is
taken into account ($4!$ permutations of the $4$ vertices of a tetrahedron), we
end up with the total of $6! / 4! = 30$ possible inequivalent tetrahedra. Those
$30$ tetrahedra have different volumes, which means that if we provide six
numbers $l_{ij}$, there is $30$ inequivalent different ways to position them in
the determinant above.

A dihedral angle of a tetrahedron is given by the formula
$$
\sin \theta_a = \frac{3}{2} \frac{a V_{abcdef}}{A_{abf}A_{ace}}.
$$
The angle $\theta_a$ corresponds to the edge $a$, and it is
constructed between the outward-normal vectors of the faces $a,b,f$ and $a,c,e$,
such that the edge $d$ does not intersect the edge $a$. This angle is equal to
$\pi -
\varphi_a$, where $\varphi_a$ is the angle between the faces $a,b,f$ and
$a,c,e$.

Regarding the above equation, once we fix the value of the right-hand
side, there are in general {\em two different angles} which satisfy the
equation. This is a consequence of double-valuedness of the arcsine function on
the $[0,\pi]$ codomain. However, when  the edge lengths are given, $\theta_a$
can be expressed as 
$$
\theta_a = \left\{
\begin{array}{ccl}
\ds \pi - \arcsin \left( \frac{3}{2} \frac{a V_{abcdef}}{A_{abf}A_{ace}} \right)
, & \text{if} & d \leq \frac{\sqrt{ a^2(b^2 + c^2 + e^2 + f^2 -
a^2) + (e^2 - c^2)(b^2 - f^2)}}{\sqrt{2}a}, \\
\ds \arcsin \left( \frac{3}{2} \frac{a V_{abcdef}}{A_{abf}A_{ace}} \right), &
\text{if} & d > \frac{\sqrt{ a^2(b^2 + c^2 + e^2 + f^2 - a^2) +
(e^2 - c^2)(b^2 - f^2)}}{\sqrt{2}a}. \\
\end{array}
\right.
$$

Finally, it is convenient to notice that the angle $\theta_a$ is unchanged if we
scale all
edge lengths of the tetrahedron, i.e. if we multiply them all with the same
positive constant. This is easily visualized since the tetrahedron does not
change its ``shape'' if we ``zoom in/out'', and can be verified analytically by
inspecting the above equation. In other words, we have an identity
$$
\theta_{\lambda a} (\lambda a, \lambda b, \lambda c, \lambda d, \lambda e,
\lambda f ) = \theta_a (a,b,c,d,e,f), \qquad \lambda >0\,.
$$

By differentiating this identity with respect to $\lambda$ and $a,b,c,d,e,f$, it
is easy to
derive the so-called {\em Schl\" afli differential identity} for a
tetrahedron
 \begin{equation} \label{SchlafliIdentitet}
\sum_{s=1}^6 l_s \frac{\del \theta_{s}}{\del l_p} = 0, \qquad \forall l_p\in \{
a,b,c,d,e,f \}\,.
 \end{equation}

As far as the $6j$ symbols are concerned, we are mainly interested in the
asymptotic formulae for
large spins. The limit of large spins is defined as a limit when the scaling
parameter
$\Lambda$ tends to infinity. The relation between a spin and $\Lambda$ is
defined by
$$
\lim_{j\to \infty} f(j) \equiv \lim_{\Lambda \to \infty} f( \Lambda l_k  -
\frac{1}{2}), \qquad \text{ where } \qquad l_k= k+\frac{1}{2}.
$$
Here $k$ is some initial finite spin, which is scaled to $j$ via $\Lambda$
according to the map
$$
j + \frac{1}{2} = \Lambda \left( k + \frac{1}{2} \right)\,.
$$
In other words, it is not spins themselves that are being scaled, but rather
their corresponding {\em edge lengths}.

The large-spin asymptotics of a $6j$ symbol can be defined in the following way.
Fix the $6$
initial spins $k_1,\dots, k_6$, and associate to them a tetrahedron with
the edge lengths $l_i=k_i+\frac{1}{2}$. This tetrahedron is then
scaled by the parameter $\Lambda$ into a new tetrahedron with
the edge lengths $\Lambda l_i$. Let us denote the corresponding spins as $j_i$,
then the
following asymptotic equation holds \cite{PR, Dupuis2009}:
 \begin{equation} \label{AsimptotikaSestJ}
\lim_{j\to\infty} \sestj{j_1}{j_2}{j_3}{j_4}{j_5}{j_6} \equiv
\lim_{\Lambda\to\infty} \sestj{\Lambda l_1 - \frac{1}{2}}{\Lambda l_2 -
\frac{1}{2}}{\Lambda l_3 - \frac{1}{2}}{\Lambda l_4 - \frac{1}{2}}{\Lambda l_5 -
\frac{1}{2}}{\Lambda l_6 - \frac{1}{2}}
=\frac{1}{\sqrt{12\pi \Lambda^3 V(l)}} \cos \cS(\Lambda,l) \,,
 \end{equation}
where
 \begin{equation} \label{PonzanoReggeDejstvoSaPopravkom}
\cS(\Lambda, l) = \Lambda \sum_{s=1}^6 l_s \theta_s(l) + \frac{\pi}{4} +
\frac{1}{\Lambda}\DL(l) + \cO\left(\frac{1}{\Lambda^2} \right)\,.
 \end{equation}
The first term in $\cS$ represents the familiar Regge action, while $\DL(l)$ is
a very complicated correction of $O(1/\Lambda)$, see Ref. \cite{Dupuis2009},
equation (32). 

The second possible configuration is when $5$ spins in a $6j$ symbol are large,
of $O(\Lambda)$, while one is of $O(1)$. In this case, we have the following
asymptotic formula
\begin{equation} \label{AsimptotikaPetJ}
\lim_{j\to\infty} \sestj{j_1}{j_2}{j_3}{j_2+k_2}{j_1+k_1}{k_3}
=
\frac{(-1)^{j_1+j_2+j_3+k_2+k_3}}{\sqrt{(2j_1+1)(2j_2+1)}}
d^{(k_3)}_{k_2,k_1}(\theta)\,,
\end{equation}
where
$$
\cos\theta = \frac{j_1(j_1+1)+ j_2(j_2+1) -
j_3(j_3+1)}{2\sqrt{j_1(j_1+1)j_2(j_2+1)}},
\qquad 0 \leq \theta \leq \pi\,.
$$
Here $d^{(k_3)}_{k_2,k_1}(\theta)$ are the usual matrix elements of the $SU(2)$
rotation operator, see \cite{PR}.

Next we have a configuration with only $4$ large spins, and the corresponding
asymptotics is given by
$$
\lim_{j\to\infty} \sestj{k_1}{j+k_2}{j+k_3}{k_4}{j+k_5}{j}
=
\frac{(-1)^{k_1+k_4+\min(k_2+k_5,k_3)}}{|k_2+k_5-k_3|!}
(2j)^{-1-|k_2+k_5-k_3|} \left[ 1+ \cO\left( \frac{1}{j^2} \right) \right]
$$
\begin{equation} \label{AsimptotikaCetiriJ}
\left[ \frac{(k_1-k_2+k_3)!(k_1-k_5)!(k_4-k_5+k_3)!(k_4-k_2)!}{
(k_1+k_2-k_3)!(k_1+k_5)!(k_4+k_5-k_3)!(k_4+k_2)!} \right]^{\frac{1}{2} \sgn
(k_3-k_2-k_5)}
\,.
\end{equation}
see \cite{PR}.

Finally, the configuration with only $3$ large spins has the asymptotic formula
proportional
to the Wigner $3j$ symbol \cite{PR}
\begin{equation} \label{AsimptotikaTriJ}
\lim_{j\to\infty} \sestj{k_1}{k_2}{k_3}{j+k_4}{j+k_5}{j}
=
(-1)^{k_1+k_2+k_3+2(k_4+k_5)} \frac{1}{\sqrt{2j}}
\trij{k_1}{k_2}{k_3}{k_5}{-k_4}{k_4-k_5}\,.
\end{equation}

As a final remark, note that the spins and the corresponding edge lengths for a
$6j$ symbol are
constrained by the triangle inequalities.
However, these constraints are {\em not sufficient} to guarantee the condition
$V^2 >0$
for the volume of the $6j$ symbol tetrahedron. There are
choices of spins such that the $6j$ symbol is well defined, and the left-hand
side
of (\ref{AsimptotikaSestJ}) is real, while $V^2<0$. Consequently the
right-hand side of (\ref{AsimptotikaSestJ}) is a complex number. Of course, in
such situations the equation
(\ref{AsimptotikaSestJ}) does not apply, and we have a different, exponentially
decreasing asymptotics, see \cite{PR}. However, we are interested in the
extremal points of the asymptotic formula, and the exponentially decreasing
asymptotics does not have any such points, so it is not relevant for our
purposes.

\section{\label{DodatakKosinusnaFormula}The cosine approximation}

In this section we shall give a rigorous derivation of the ``cosine
formula'' (\ref{AproksimacijaKosinusneFunkcije}). We start by introducing the
Heaviside step-function as
$$
H(x) = \left\{
\begin{array}{ccl}
0 & & \text{ for } x<0, \\
\frac{1}{2} & & \text{ for } x=0, \\
1 & & \text{ for } x>0. \\
\end{array}
\right.
$$

This is a standard definition, and the standard rules apply. For example,
derivative of a
Heaviside function is the Dirac $\delta$ function and the integral
of $\delta$-function from $-\infty$ to $x$ gives $H(x)$. We
can then define the so-called rectangle function
$$
K_a(x) \equiv H(x+a) H(-x+a) = \left\{
\begin{array}{ccl}
0 & & \text{ for } x<-a\,, \\
\frac{1}{2} & & \text{ for } x=-a\,, \\
1 & & \text{ for } -a<x<a, \\
\frac{1}{2} & & \text{ for } x=a\,, \\
0 & & \text{ for } x>a\,. \\
\end{array}
\right.
$$

The function $K_a(x)$ is equal to one in the interval $(-a,a)$,
zero outside, and at the points $-a$ and $a$ $K$ is conveniently defined to be
$\frac{1}{2}$, so that it has a nice property proven in Lemma 1.

\medskip

\textbf{Lemma 1.} The following identities hold
 \begin{equation}
K_b(x+(a+b)) + K_a(x) + K_b(x-(a+b)) = K_{a+2b}(x)\,,\label{kid1}
 \end{equation}
 \begin{equation}
\sum_{k\in \celi} K_a(x-2ka) = 1\,.
 \label{kid2}\end{equation}

\textit{Proof}. The first identity can be demonstrated by using the definition
of $K$. First,
we see that outside of the interval $[-a-2b,a+2b]$ both the left-hand and the
right-hand
sides of (\ref{kid1}) are equal to zero. Next, the three terms on the left-hand
side of (\ref{kid1}) are equal
to one respectively in the intervals $(-a-2b,-a)$, $(-a,a)$ and $(a,a+2b)$.
Finally, at points $-a$ and $a$ the appropriate terms are equal to $\frac{1}{2}$
and
add up to one so that the resulting function is continuous and equal to
one in the whole interval $(-a-2b,a+2b)$. At the boundary points of
$[-a-2b,a+2b]$,
the left-hand side of (\ref{kid1}) gives the contribution of $\frac{1}{2}$. All
these results
taken together form by definition the right-hand side of (\ref{kid1}). 

The identity (\ref{kid2}) can be proved by applying the identity (\ref{kid1})
iteratively. First
notice that the sum over all integers is in fact defined as the limit
$m\to \infty$ of the sum over the domain $-m, -m+1, \dots, m-1,m$. Consequently
we can write:
$$
\sum_{k\in\celi} K_a(x-2ka) = \lim_{m\to\infty} \sum_{k=-m}^m K_a(x-2ka)\,.
$$

Now note that by applying the first identity $m$ times we have
$$
\sum_{k=-m}^m K_a(x-2ka) = K_{(2m+1)a}(x)\,.
$$
This identity can be also easily seen graphically, and can be proved by
induction over
$m$. Therefore,
$$
\sum_{k\in\celi} K_a(x-2ka) = \lim_{m\to\infty} K_{(2m+1)a}(x) = K_{\infty}(x)
=1\,.
$$
\textit{End of proof}.

\medskip

The function $K_a(x)$ was introduced because it allows for a neat ``cutting'' of
the appropriate pieces of the real line. Now we make use of this property in
order to prove the
following identity.

\medskip

\textbf{Lemma 2.} The identity
$$
\cos x = \sum_{k\in\celi} (-1)^k K_{\frac{\pi}{2}}(x-k\pi) \cos(x-k\pi)
$$
holds.

\medskip

\textit{Proof}. Start from the right-hand side and compute the left-hand side
in the following way
$$
\begin{array}{ccl}
{\rm RHS} & = & \ds\sum_{k\in\celi} (-1)^k K_{\frac{\pi}{2}}(x-k\pi) \left( \cos
x \cos k\pi + \sin x\sin k\pi \right) \\
         & = & \ds\sum_{k\in\celi} (-1)^k K_{\frac{\pi}{2}}(x-k\pi) (-1)^k \cos
x \\
         & = & \ds\cos x \sum_{k\in\celi} K_{\frac{\pi}{2}}(x-k\pi) \\
         & = & \ds\cos x \,. \\
\end{array}
$$
\textit{End of proof}.

\medskip

\noindent Now if we note that $K_{\frac{\pi}{2}}(x-k\pi) \cos(x-k\pi)$ is
continuous and non-negative for all $x\in\realni$ and $k\in\celi$, the statement
of Lemma 2 can be rewritten in the form
 \begin{equation}
\cos x = \sum_{k\in\celi} (-1)^k e^{\ln \left[ K_{\frac{\pi}{2}}(x-k\pi)
\cos(x-k\pi) \right] }\,,
 \end{equation}
which represents the key ``cosine formula''. The exponent can be expanded into a
power series around the point $x=k\pi$,
$$
\ln \left[ K_{\frac{\pi}{2}}(x-k\pi) \cos(x-k\pi) \right] =
-\frac{1}{2}(x-k\pi)^2 - \frac{1}{12}(x-k\pi)^4 + \cO(x-k\pi)^6, \qquad (x\to
k\pi)\,,
$$
and in the leading order approximation we can write
$$
\cos x \approx \sum_{k\in\celi} (-1)^k e^{ -\frac{1}{2}(x-k\pi)^2 + R(x,k) }\,,
$$
where $R(x,k)$ is the remainder of $O((x-k\pi)^4)$ when $x\to k\pi$.

This kind of approximation is useful since the cosine function is well
approximated in the vicinity of {\em all extremal points simultaneously}. In
the crudest approximation, the remainder $R(x,k)$ can be substituted by some
average value over one period of the cosine, $\bar{R}$, so we write:
$$
\cos x \approx e^{\bar{R}} \sum_{k\in\celi} (-1)^k e^{ -\frac{1}{2}(x-k\pi)^2},
\qquad x\to k\pi.
$$

The constant $\bar R$ can be calculated, because at any point $x=k_0\pi$
($k_0\in\celi$) the formula must become {\em exact}, with no error. Namely,
expanding the exact exponent into a power series and discarding certain terms we
effectively allow for the appearance of the ``tails'' in each Gaussian in the
sum. These tails then give artificial contribution to other Gaussians, and the
constant $e^{\bar{R}}$ accounts for the appropriate ``correction''. For
example, substituting $x=0$ into the left and right side, we have
$$
1 = e^{\bar{R}} \sum_{k\in\celi} (-1)^k e^{ \frac{\pi^2}{2}k^2}\,,
$$
so that
$$
e^{\bar{R}} = \frac{1}{\vartheta_4(0,e^{-\frac{\pi^2}{2}})} \approx 1,01459\,.
$$

Here $\vartheta_4(u,q)$ is the so-called inverse elliptic theta-function of the
fourth kind. Of course, if we had kept the $x^4$ term in the exponent, the
Gaussian tails would have been different and the constant $\bar R$ would have
had different
numerical value, but again such that the equation is exact at all points
$x=k\pi$. If we had kept all terms in the exponent, the tails would have
vanished
and this constant would have been equal to one.

Anyway, we are interested only in the crudest approximation of the
cosine function with Gaussian functions, so that the formula we will use
is
 \begin{equation} \label{KosinusnaFormula}
\cos x \approx \frac{1}{\vartheta_4(0,e^{-\frac{\pi^2}{2}})} \sum_{k\in\celi}
(-1)^k e^{ -\frac{1}{2}(x-k\pi)^2}, \qquad x\to k\pi \,.
 \end{equation}

One very important remark here is that this equation makes sense {\em only in
the vicinity of cosine extremal points}, i.e. in the neighborhood of the points
$x=k\pi$. Therefore one should make sure that the argument of
the cosine is close enough to {\em some} integer multiple of $\pi$, whenever
one applies the formula. For the points in the vicinity of $k\pi/2$, the
approximation is less accurate.

\section{\label{SaddlePoint}Laplace's method}

Let $f(x)$ be a real function in the interval $[a,b]$ such that it has a single
global maximum at a point $x_0 \in [a,b]$. We would like to find the asymptotics
of the integral
$$
I= \int_a^b dx \; e^{\Lambda f(x)} \,,
$$
for large $\Lambda$. We will describe here the Laplace method of calculating
this asymptotics.

The idea of the method is the fact that $e^{\Lambda f(x)}$ goes to
infinity at the fastest rate at the point of the global maximum as
$\Lambda\to\infty$. One can then  approximate $e^{\Lambda f(x)}$ in $[a,b]$ by a
Gaussian function centered at $x_0$. This also implies that the integration
domain $[a,b]$ can be
extended to the whole real line, since the ``tails'' of a Gaussian function do
not give
a significant contribution. By using the Taylor series for $f(x)$ at the point
$x_0$
we obtain 
$$
f(x) = \sum_{n=0}^\infty {f^{(n)}(x_0)\over n!}(x-x_0)^n = f(x_0) - \frac{1}{2}
|f''(x_0)| (x-x_0)^2 + O(x-x_0)^3 \,.
$$

By changing the integration variable to $y=x-x_0$, the integral $I$ is reduced
to a sum of
Gaussian integrals
$$
I= e^{\Lambda f(x_0)} \int_{\realni}
dy\,  e^{-\frac{\Lambda}{2}|f''(x_0)| y^2} \left(\sum_{n=0}^{\infty}
\frac{\Lambda^n}{n!}\left[ O(y^3) \right]^n \right)\,
$$
which implies
\begin{equation}
I = e^{\Lambda f(x_0)}
\sqrt{\frac{2\pi}{\Lambda|f''(x_0)|}} \left[ 1 +
O\left(\frac{1}{\Lambda}\right)\right] \,.\label{lap}
\end{equation}

In the case when there are several global maxima in $[a,b]$ with the same value
of $f(x)$, then the formula (\ref{lap}) becomes
\begin{equation}
I = \sum_{x_0 \in [a,b]} e^{\Lambda f(x_0)}
\sqrt{\frac{2\pi}{\Lambda|f''(x_0)|}} \left[ 1 +
O\left(\frac{1}{\Lambda}\right)\right] \,.\label{flap}
\end{equation}

\section{\label{DodatakGausoviIntegrali}Generalized Gaussian integrals}

Let us start from the identity
$$
\int_{\realni} dx\, e^{-\frac{1}{2}ax^2 + bx} = \sqrt{\frac{2\pi}{a}}\,
e^{\frac{b^2}{2a}} \,,\,\, a>0 \,.
$$
This identity can be generalized to $\realni^n$ 
\begin{equation}
\int_{\realni^n}d^nx \, e^{-\frac{1}{2}x^T A x + Bx} =
\sqrt{\frac{(2\pi)^n}{\det A}}\, e^{\frac{1}{2} BAB^T} \,,\label{ndgi}
\end{equation}
where $A$ and $B$ are matrices of the type $n\times n$ and $1\times n$
respectively
and $A$ is a symmetric matrix. This identity can be derived by
reducing the integral in (\ref{ndgi}) to a product of one-dimensional integrals
by a change of variables $x'=Ox$, where
$O$ is the orthogonal matrix which diagonalizes the matrix $A$. In the diagonal
basis the
integral reduces to a product of $n$ one-dimensional integrals. The critical
assumption here is that the eigenvalues of $A$ must be strictly positive if the
integral is to converge. However, we are interested in the situation when some
of the
eigenvalues of $A$ are zero. In this case the integral  in (\ref{ndgi})
diverges, and
our goal now is to regularize this divergence in one special case.

Let us start with an integral over a compact domain $D = [-G,G]^n$
$$
I_D = \int_D d^nx \, e^{-\frac{1}{2}x^T A x + Bx}.
$$
For simplicity, assume that the matrix $A$ is already diagonal, and denote its
rank as $r<n$. Also, denote all $n$ eigenvalues as $a_i$, and let $a_1,\dots,
a_r \neq 0$, while $a_{r+1},\dots,a_n = 0$. Next, let $K$ be the projector to
the null-space of matrix $A$, and assume that $BK=0$. This means that in this
particular basis we have $B_{r+1},\dots,B_n =0$. Now the integral $I_D$ can be
split into a product of $r$ Gaussian one-dimensional integrals and $n-r$
one-dimensional integrals of a constant function, where the constant is equal to
one, 
$$ I_D = \prod_{i=1}^r \int_{-G}^G dx_i \, e^{-\frac{1}{2}x_i^2 a_i + B_ix_i}
\prod_{i=r+1}^n \int_{-G}^G dx_i \, \underbrace{e^{-\frac{1}{2}x_i^2 a_i +
B_ix_i}}_{1}\,.$$

In the limit $G\to\infty$ we will define a regularized $I$
\begin{equation} \label{GeneralisaniGausovIntegral}
 I = {\lim_{G\to\infty} I_D \over  \lim_{G\to\infty} \left[\int_{-G}^G dx 
\right]^{n-r}}= \sqrt{\frac{(2\pi)^r}{\det M}}\, e^{\frac{1}{2}
NM^{-1} N^T} \,.
\end{equation}
Here $M$ and $N$ denote the submatrices of $A$ and $B$ of type $r\times r$ and
$1\times r$, which are obtained by simultaneous change of basis which puts $A$
and $B$ in block-diagonal form:
$$
A = \left[
\begin{array}{cc}
M & 0 \\
0 & 0 \\
\end{array}
\right], \qquad B = \left[
\begin{array}{cc}
N & 0 \\
\end{array}
\right] .
$$
Simultaneous diagonalization is possible because of the imposed assumption
$BK=0$.

\section{\label{DodatakMatricneTeoreme}Matrix theorems}

Here we explain some results for matrices that we have used in the main text.
These results
can be found in \cite{Zhang2005}. However, one of the results, the statement (c)
bellow, is a new result, to the best of our
knowledge.

\medskip

\textbf{Theorem 1.} Let $\Delta$ be a symmetric real matrix of type $n\times n$
 and let $R$ be its rank. Let us split $\Delta$ into blocks as
$$
\Delta = \left[
\begin{array}{cc}
S & N \\
N^T & M \\
\end{array}
\right],
$$
where $S$ is a $J\times J$ matrix, $N$ is a $J\times r$ matrix, $M$ is a
$r\times r$ matrix  and $n = J + r$. We will also assume that $M$ is invertible.

Let us construct the Schur complement (see \cite{Zhang2005}) $\tS$, which is a
$J\times J$ matrix
$$
\tS = S - N M^{-1} N^T.
$$
Denote the rank of $\tS$ as $\rho$. Then
\begin{itemize}
\item[(a)] $R = r + \rho$ (Guttman rank additivity);
\item[(b)] $\det \Delta = \det \tS \det M$ (Schur determinant formula);
\item[(c)] if $0< \rho < J$, then
 \begin{equation} \label{IdentitetZaDeterminante}
\det M_{\Delta} (\det B_4)^2 = \det M \det M_{\tS}.
 \end{equation}
\end{itemize}

Here $M_{\Delta}$ and $M_{\tS}$ are invertible $R\times R$ and $\rho\times\rho$
matrices,
 respectively. They are obtained by using orthogonal
transformations which put $\Delta$ and $\tS$ into a block-diagonal form
$$
\Delta = \left[
\begin{array}{cc}
0 & 0 \\
0 & M_{\Delta} \\
\end{array}
\right], \qquad
\tS = \left[
\begin{array}{cc}
0 & 0 \\
0 & M_{\tS} \\
\end{array}
\right],
$$
The $B_4$ matrix will be explicitly constructed in the proof below.

\medskip

\textit{Proof}. We start from the Aitken block diagonalization
formula \cite{Zhang2005} and from now on we use $I$ to denote a unit matrix of
any size
appropriate for its position in an equation
 \begin{equation} \label{AitkenFormula}
\left[
\begin{array}{cc}
I & -NM^{-1} \\
0 & I \\
\end{array}
\right]
\left[
\begin{array}{cc}
S & N \\
N^T & M \\
\end{array}
\right]
\left[
\begin{array}{cc}
I & 0 \\
-M^{-1}N^T & I \\
\end{array}
\right]
=
\left[
\begin{array}{cc}
\tS & 0 \\
0 & M \\
\end{array}
\right] \,.
 \end{equation}

This equation can be verified by a direct multiplication of the left-hand side.
Denoting
the first matrix on the left as $A$, we can rewrite this identity in a compact
form $A\Delta A^T = \tS \oplus M$. The rank of the right-hand side is the sum
of ranks of $\tS$ and $M$, which amounts to $\rho+r$. Since the rank of $A$ is
equal to its dimension $n$, the total rank of the product on the left-hand side
is equal to the rank of $\Delta$, so we easily obtain
$$
R = r + \rho\,,
$$
which completes the proof of part (a).

Next, we take the determinant of (\ref{AitkenFormula}). Since $A$ is
block-triangular, its determinant is a product of determinants of blocks on
the diagonal, so that we obtain $\det A =1$. The left-hand side is thus the
product of
determinants, $\det A \det \Delta \det A^T$, and it is equal to $\det\Delta$
because $\det A^T=\det A =1$. On the right-hand side we have a block-diagonal
matrix, so that its determinant is equal to $\det\tS \det M$. Hence,
$$
\det \Delta = \det\tS \det M \,,
$$
which completes the proof of part (b).

In order to prove (c), let $O$ be a $J\times J$ orthogonal matrix
which transforms $\tS$ into a block-reduced form,
$$
O\tS O^T = 0\oplus M_{\tS}\,.
$$
Since $\rho\neq 0$, matrix $\tS$ has exactly $\rho$ nonzero eigenvalues, which
constitute $M_{\tS}$, and since $\tS$ is also real and symmetric, there will
always
exist an orthogonal matrix $O$ that diagonalizes it. Given that the
eigenvalues of $M_{\tS}$ are nonzero, it is invertible. The zero-block is of
type $\nu\times\nu$, where $\nu = J-\rho$ is the dimension of the null-space of
$\tS$. By using $O$ one can construct an  orthogonal $n\times n$ matrix $P =
O\oplus I$ such that
 \begin{equation} \label{MatricaP}
P \left( \tS \oplus M \right) P^T = 0\oplus M_{\tS} \oplus M.
 \end{equation}

By using an analogous argument one can always construct an orthogonal $n\times
n$ matrix $Q^T$ such that
 \begin{equation} \label{MatricaQ}
Q^T\Delta Q = 0\oplus M_{\Delta}, \qquad \Leftrightarrow \qquad
\Delta = Q \left( 0\oplus M_{\Delta} \right) Q^T \,.
 \end{equation}
The zero block comes from the null-space of $\Delta$. It is of the size $n-R$,
which is also equal to $\nu$, since $n=J+r$ and $R = r+\rho$ according to the
part
(a) of Theorem 1.

Consider (\ref{AitkenFormula}), and multiply it by $P$ from the left and by
$P^T$ from the right, and use (\ref{MatricaP}) and (\ref{MatricaQ}) to rewrite
it in the form
 \begin{equation} \label{VezaMinoraZaDeltaIsTilda}
PAQ \left( 0\oplus M_{\Delta} \right) Q^T A^T P^T = 0\oplus M_{\tS} \oplus M \,.
 \end{equation}

Let us introduce the matrix $B\equiv PAQ$ and write it in the block form as
$$
B = \left[
\begin{array}{cc}
B_1 & B_2 \\
B_3 & B_4 \\
\end{array}
\right],
$$
where the blocks $B_1$, $B_2$, $B_3$ and $B_4$ are $\nu\times\nu$,
$\nu\times R$, $R\times\nu$ and $R\times R$ matrices, respectively. Substituting
this
into the left-hand side of (\ref{VezaMinoraZaDeltaIsTilda})
yields
\begin{equation} \label{LevaStranaMatricneJednacine}
PAQ \left( 0\oplus M_{\Delta} \right) Q^T A^T P^T \equiv
B \left[
\begin{array}{cc}
0 & 0 \\
0 & M_{\Delta} \\
\end{array}
\right] B^T = \left[
\begin{array}{cc}
B_2 M_{\Delta} B_2^T & B_2 M_{\Delta} B_4^T \\
B_4 M_{\Delta} B_2^T & B_4 M_{\Delta} B_4^T \\
\end{array}
\right].
\end{equation}

By comparing (\ref{LevaStranaMatricneJednacine}) to the right-hand side of
(\ref{VezaMinoraZaDeltaIsTilda}), we obtain
\begin{equation} \label{IzjednacenaLevaIdesnaStrana}
\left[
\begin{array}{cc}
B_2 M_{\Delta} B_2^T & B_2 M_{\Delta} B_4^T \\
B_4 M_{\Delta} B_2^T & B_4 M_{\Delta} B_4^T \\
\end{array}
\right]
=
\left[
\begin{array}{ccc}
0 & 0 & 0 \\
0 & M_{\tS} & 0 \\
0 & 0 & M \\
\end{array}
\right]\,.
\end{equation}

Note that the zero-block of (\ref{IzjednacenaLevaIdesnaStrana}) is a
$\nu\times\nu$ matrix, which is also the $B_2
M_{\Delta} B_2^T$ block. We then read off the following equations
 \begin{equation} \label{VezaMovaIbeCetiri}
B_4 M_{\Delta} B_4^T = M_{\tS} \oplus M\,,
 \end{equation}
 \begin{equation} \label{JednacinaZaBeDva}
B_2 M_{\Delta} B_4^T = 0\,,
 \end{equation}
 \begin{equation} \label{DrugaJednacinaZaBeDva}
B_2 M_{\Delta} B_2^T = 0\,.
 \end{equation}

By taking the determinant of (\ref{VezaMovaIbeCetiri}), we finally obtain
$$
\det M_{\Delta} (\det B_4)^2 = \det M \det M_{\tS} \,.
$$
This establishes (\ref{IdentitetZaDeterminante}) and completes the proof of
part (c) of the theorem.

Given that $M$, $M_{\tS}$ and $M_{\Delta}$ are all invertible, we have $\det
B_4 \neq 0$ which means that $B_4$ is also invertible. By multiplying
(\ref{JednacinaZaBeDva}) by $(B_4^T)^{-1}M_{\Delta}^{-1}$ from the right, we
obtain
$$
B_2 = 0.
$$
The equation (\ref{DrugaJednacinaZaBeDva}) now vanishes and does not provide
any additional constraint. Therefore, the matrix $B$ has the following form
 \begin{equation} \label{OblikMatriceB}
B \equiv PAQ = \left[
\begin{array}{cc}
B_1 & 0 \\
B_3 & B_4 \\
\end{array}
\right] \,.
 \end{equation}
\textit{End of proof}.

\medskip

\textbf{Remark 1.} The $\Delta$ matrix from the main text has the form
$$
\Delta = \left[
\begin{array}{ccc}
S & N & 0 \\
N^T & M & 0 \\
0 & 0 & 0 \\
\end{array}
\right],
$$
which differs from the one in Theorem 1 by an additional zero-block of size
$\Omega-n$. However, these additional zeroes are integrated out before the
Theorem 1 is applied, and they do not affect the statements of Theorem 1.

\medskip

\textbf{Remark 2.} The result (c) is a generalization of the result (b) to the
case
when $\Delta$ is a singular matrix. While the part (b) is in fact valid for
singular matrices, it
merely states that $0=0$ and provides no information about nonsingular principal
minors of $\Delta$. The result (c) is more fine-grained, and provides precisely
this nontrivial information about $\Delta$.

It was assumed in the part (c) that $0<\rho < J$. If $\rho = J$ then $\Delta$ is
a regular matrix, and hence the result (b) can be used. If $\rho = 0$, then $\tS
= 0$, $\nu = J$,
and instead of (\ref{VezaMovaIbeCetiri}) we obtain
$$
B_4 M_{\Delta} B_4^T = M \,,
$$
and consequently
$$
\det M_{\Delta} (\det B_4)^2 = \det M\, .
$$

In this case we can set $P=I$ and obtain
$$
B \equiv AQ = \left[
\begin{array}{cc}
B_1 & 0 \\
B_3 & B_4 \\
\end{array}
\right]
$$
for the matrix $B$.

\medskip

\textbf{Remark 3.} In the main text we use the results (b) and (c) to determine
the
leading $\Lambda$-order of the Schur complement $\tS$, knowing that $\Delta$ is
of $O(1)$. However, it is necessary to show that $B_4$ is of
$O(1)$ as well. In order to do this, note that
$$
\det B = \det P \det A \det Q = \pm 1 \,,
$$
since $P$ and $Q$ are unitary matrices. On the other hand, from
(\ref{OblikMatriceB}) we know that $\det B = \det B_1 \det B_4$, so that we have
 \begin{equation} \label{VezaDeterminantiBjedanIBcetiri}
\det B_1 \det B_4 = \pm 1 \,.
 \end{equation}

Let us now assume that the blocks $B_1$ and $B_4$ are of order $k$ and $m$ in
$1/\Lambda$, respectively
$$
B_1 = \frac{C}{\Lambda^k} + \cO\left( \frac{1}{\Lambda^{k+1}} \right), \qquad
B_4 = \frac{D}{\Lambda^m} + \cO\left( \frac{1}{\Lambda^{m+1}} \right), \qquad
k,m\geq 0, \qquad C,D \sim O(1) \,.
$$
The numbers $k$ and $m$ cannot be negative since the whole $B$ matrix must be of
$O(1)$. Namely, the matrices $P$ and $Q$ are orthogonal, and consequently
all their elements are bounded above by $1$. Thus $P$ and $Q$ are of $O(1)$.
The matrix $A$ is also of $O(1)$, since $\Delta$ and consequently $M,N,M^{-1}$
are
all of the same order. Therefore, $B = PAQ \sim O(1)$.

Since $B_1$ is a $\nu\times\nu$ matrix and $B_4$ is a $R\times R$ matrix,
then
\begin{equation} \label{DeterminanteBjedanIBcetiri}
\det B_1 = \frac{1}{\Lambda^{k\nu}} \det C + \cO\left( \frac{1}{\Lambda^{k+1}}
\right)\,, \qquad
\det B_4 = \frac{1}{\Lambda^{mR}} \det D + \cO\left( \frac{1}{\Lambda^{m+1}}
\right)\,.
\end{equation}

By substituting (\ref{DeterminanteBjedanIBcetiri}) back into
(\ref{VezaDeterminantiBjedanIBcetiri}) we obtain the
consistency equation
$$
k\nu + mR = 0 \,.
$$
Since both $\nu,R > 0$ while $k,m\geq 0$, the only solution of this equation is
$k=m=0$. Therefore
$$
\det B_4 \sim B_4 \sim O(1) \,.
$$

In the case when $\nu=0$ the $\Delta$ matrix is regular and instead of the part
(c) we use the part (b) of Theorem 1. However, the part (b) does not involve
$\det B_4$, so that we need the above result only for $\nu>0$.

\bibliographystyle{my-h-elsevier}

\begin{thebibliography}{10}

\bibitem{lqg}
C. Rovelli,
Quantum Gravity,
Cambridge University Press (2004).

\bibitem{mik1}
A. Mikovi\'c,
Quantum gravity vacuum and invariants of embedded spin networks,
Class. Quant. Grav. {\bf 20} (2003) 3483.

\bibitem{mik2}
A. Mikovi\'c,
Flat spacetime vacuum in loop quantum gravity,
Class. Quant. Grav. {\bf 21} (2004) 3909,
Errata: Class. Quant. Grav. {\bf 23} (2006) 5459.

\bibitem{marm}
J. F. Martins and A. Mikovi\'c,
Invariants of spin networks embedded in three-manifolds,
Commun. Math. Phys. {\bf 279} (2008) 381.

\bibitem{mik3}
A. Mikovi\'c,
Spin network wavefunction and nonperturbative graviton propagator,
Fortschr. Phys. {\bf 56} (2008) 475.

\bibitem{ro2}
C. Rovelli,
Graviton propagator from background-independent quantum gravity,
Phys. Rev. Lett. {\bf 97} (2006) 151301.

\bibitem{ro3}
E. Bianchi, L. Modesto, C. Rovelli and S. Speziale,
Graviton propagator in loop quantum gravity,
Class. Quant. Grav. {\bf 23} (2006) 6989.

\bibitem{T} V. Turaev, Quantum Invariants of knots and three-manifolds, de Gruyter Studies
in Mathematics, 18,
Walter de Gruyter \& Co., Berlin, 1994

\bibitem{PR}
G. Ponzano and T. Regge,
Semiclassical limit of Racah coefficients,
in Spectroscopic and Group Theoretical Methods in Physics, edited by F. Block,
North Holland, Amsterdam (1968).

\bibitem{Dupuis2009}
M. Dupuis and E. Livine,
Pushing further the asymptotics of the 6j-symbol,
Phys. Rev. D {\bf 80} (2009) 024035.

\bibitem{Zhang2005}
F. Zhang,
The Schur Complement and its Applications,
Springer, New York (2005).

\bibitem{epr}
J. Engle, R. Pereira and C. Rovelli,
The loop-quantum-gravity vertex-amplitude,
Phys. Rev. Lett. {\bf 99} (2007) 161301.

\bibitem{elpr}
J. Engle, E. Livine, R. Pereira and C. Rovelli,
LQG vertex with finite Immirzi parameter,
Nucl. Phys. {\bf B799} (2008) 136-149.

\bibitem{fk}
L. Freidel and K. Krasnov,
A New Spin Foam Model for 4d Gravity,
Class. Quant. Grav. {\bf 25} (2008) 125018.

\bibitem{fl}
L. Freidel and D. Louapre,
Diffeomorphisms and spin foam models,
Nucl. Phys. {\bf B662} (2003) 279-298.

\bibitem{Carter1995}
J. S. Catrer, D. E. Flath and M. Saito,
The Classical and Quantum 6j-symbols,
Princeton University Press, New Jersey (1995).

\end{thebibliography}

\end{document}